\documentclass[3p]{elsarticle}
\usepackage{graphicx}
\usepackage[hidelinks]{hyperref}
\makeatletter\g@addto@macro\@floatboxreset\centering\makeatother

\usepackage[T1]{fontenc}
\usepackage{amsmath}
\usepackage{booktabs}
\usepackage{placeins}
\newcommand{\Description}[1]{}

\usepackage{tikz}

\usetikzlibrary{arrows.meta}

\definecolor{lemon}{HTML}{FBF2C0}
\definecolor{taupe}{HTML}{48392A}
\definecolor{coffee}{HTML}{43281C}
\definecolor{cinnamon}{HTML}{C06E52}
\definecolor{coral}{HTML}{F96F5D}

\tikzset{rlab/.style={inner sep=0, text=taupe, font=\fontfamily{phv}\fontsize{7}{8.6}\selectfont},
  rttl/.style={inner sep=0, text=coffee, font=\fontfamily{phv}\fontsize{8}{9.6}\selectfont\bfseries},
  rtick/.style={inner sep=0, text=taupe!85, font=\fontfamily{phv}\fontsize{6.4}{7.6}\selectfont},
  rval/.style={inner sep=0, font=\fontfamily{phv}\fontsize{6.2}{7.4}\selectfont},
  rgrid/.style={draw=taupe!22, line width=0.3pt}, raxis/.style={draw=taupe!70, line width=0.4pt}}

\begin{document}

\definecolor{apivory}{HTML}{F7F5EF}
\definecolor{apnearblack}{HTML}{0F1715}
\definecolor{apcharcoal}{HTML}{293532}
\definecolor{apgreen}{HTML}{1F9D7A}
\definecolor{apmint}{HTML}{6FE3C1}
\noindent\begin{tikzpicture}
\node[fill=apivory, rounded corners=10pt, inner xsep=20pt, inner ysep=18pt, outer sep=0pt] {%
\begin{minipage}{\dimexpr\textwidth-40pt\relax}
\setlength{\parindent}{0pt}
{\raggedright\fontfamily{phv}\bfseries\fontsize{17}{21}\selectfont\color{apnearblack}
Do Language Models Still Need the Compiler?\par}
\vspace{4pt}
{\raggedright\fontfamily{phv}\bfseries\fontsize{11.5}{14.5}\selectfont\color{apcharcoal}
Direct Editing of Executables and the Decoding Bottleneck\par}
\vspace{9pt}
{\fontfamily{phv}\fontsize{10.5}{13.5}\selectfont\color{apnearblack} Camilo Chac\'on Sartori\par}
\vspace{2pt}
{\fontfamily{phv}\fontsize{9}{12}\selectfont\color{apcharcoal} Apeiron Intelligence, Spain\quad\textperiodcentered\quad\href{mailto:camilo.chacon@apeironagents.tech}{\textcolor{apgreen}{camilo.chacon@apeironagents.tech}}\par}
\vspace{12pt}
{\fontsize{9}{11.6}\selectfont\color{apnearblack}
\textbf{Context:} Software must sometimes be changed without its source code, which is lost, cannot be rebuilt or belongs to a third party. Maintenance then means patching the executable, and a language model could make that edit itself.

\vspace{3.5pt}
\textbf{Objectives:} To measure how often language models change the behaviour of a small executable by editing its bits, with and without source code and a disassembler listing, and where failed runs fail.

\vspace{3.5pt}
\textbf{Methods:} Six models receive an AArch64 executable of about 1.5 kilobytes, written as bits, and a short change request. Each answers once, without tools and under an output limit, with edits on the bits; hidden tests run the edited file. Over 100 tasks and 6,000 runs, the prompt varies in source information and in whether it adds a disassembler listing.

\vspace{3.5pt}
\textbf{Results:} With the bits only, the models solve 5 to 98 percent. The listing raises the runs solved for five models (McNemar, adjusted $p < 0.001$) and takes DeepSeek-V4.1-Flash from 48 to 98 percent; the other four solve at most 38 percent with it. Claude Opus 5.5 solves 98 percent with the bits only, and ten of its 11 failures are one bit off the compiled change. The pattern holds with no source code. Source information changes the result for two models, both with the bits only. With the bits only, five models solve data edits far more often than instruction edits.

\vspace{3.5pt}
\textbf{Conclusion:} Whether a model still needs the compiler depends on the model: one edits these executables with the bits alone, one needs only a disassembler, and four fail most runs even with it. Two bottlenecks appear: decoding the bits into instructions and producing a correct edit once they are given. For patching without source, a disassembler should run before the model. Tasks, scoring and runs are released.

\vspace{7pt}
{\fontfamily{phv}\bfseries\fontsize{8.5}{11.6}\selectfont\color{apcharcoal}Keywords:} large language models, software maintenance, binary patching, executable editing, machine code, empirical study.\par}
\vspace{11pt}
{\color{apmint}\rule{\linewidth}{0.8pt}}\par
\vspace{9pt}
\begin{minipage}[c]{0.30\linewidth}
\includegraphics[height=9.5mm]{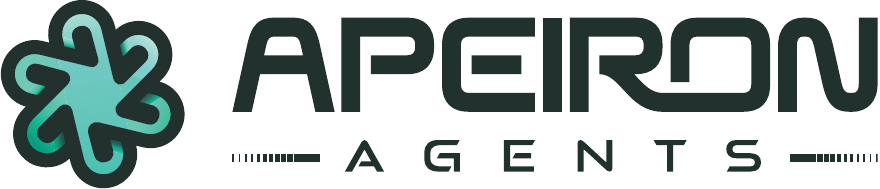}
\end{minipage}\hfill
\begin{minipage}[c]{0.66\linewidth}
\raggedleft\fontfamily{phv}\fontsize{8}{11}\selectfont\color{apcharcoal}
\textbf{Tasks, runs and code}\\
\href{https://doi.org/10.5281/zenodo.24449739}{\textcolor{apgreen}{https://doi.org/10.5281/zenodo.24449739}}
\end{minipage}
\end{minipage}};
\end{tikzpicture}

\vspace{6pt}

\section{Introduction}

Programs are usually changed through their source code. A developer edits the source and a compiler produces the new executable. Sometimes the executable is all that exists: the source is lost, the build cannot be reproduced, or the program comes from a third party. Changing a program without its source code is the subject of binary rewriting \citep{wenzl2019binaryrewriting}. Patching then works on the executable itself, for instance by decompiling it, patching the decompiled code and recompiling it \citep{dai2026scribe}. This paper asks a question of software understanding and reengineering at the level of the executable: can a language model change a compiled program by editing its bytes, with no compiler in the loop?

\begin{figure}[t]
  \centering
  \includegraphics[width=\linewidth]{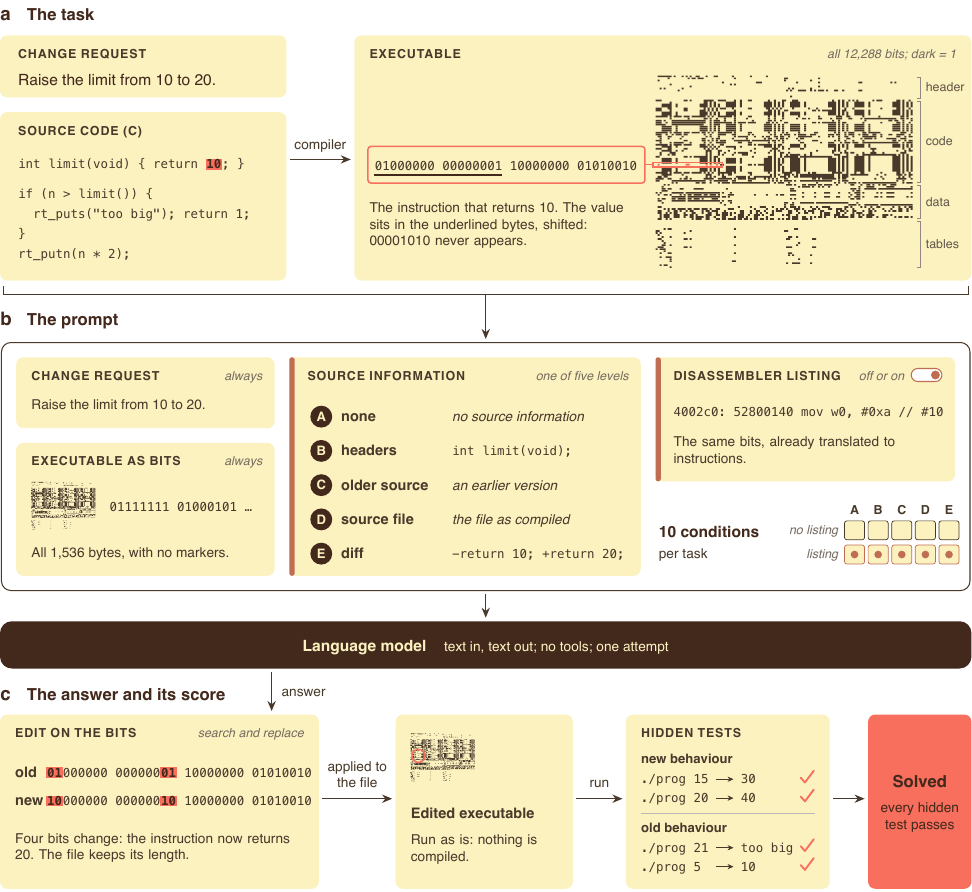}
  \caption{The design of the study on one toy task. (a) A change request, the C source and the compiled executable, drawn with one mark per 1-bit; the instruction that holds the constant is highlighted. (b) The prompt: the change request and the executable as bits are always given, the source information takes one of five levels, A to E, and the disassembler listing is absent or pasted as text, which gives 10 conditions per task. (c) The model answers with search-and-replace edits on the bits, and a run is solved when the edited file passes every hidden test.}
  \Description{Three panels. Panel a shows a change request, C source code and the executable as a block of bits. Panel b shows the parts of the prompt and a grid of ten conditions. Panel c shows the language model, its edit on the bits, the edited executable and the hidden tests.}
  \label{fig:method}
\end{figure}

This study tests that case, which we call \textit{direct editing}. Figure~\ref{fig:method} shows the design on one toy task. The model receives a small executable written out as bits, 8 per byte, and a request of one or two sentences for a change in behaviour. No tools are available. The answer is a list of search-and-replace edits on the bits, and hidden tests run the edited file as it is. We evaluate six models on 100 tasks in 20 families, which gives 6,000 runs.

The prompt varies in two ways. The first is how much of the source code it includes, in five levels from none to the exact change. The second is whether it includes a disassembler listing, which shows the same bytes as instructions with their addresses and the sections of the file. We say \emph{with the listing} for the runs whose prompt includes it and \emph{with the bits only} for the runs whose prompt does not, at any level of source information. Each run is scored by the four checks of Section~\ref{sec:scoring}, applied in order, and a failed test is attributed to locating the code, writing the new instruction, or preserving the rest of the program. The three research questions of Section~\ref{sec:design} concern source information and the listing, the kind of change, and the step at which a run fails.

For five of the six models the listing raises the runs solved, and one of them, DeepSeek-V4.1-Flash, solves nearly all its runs with it. The results separate two bottlenecks. The first, which we call the decoding bottleneck, is turning the bits into instructions. We measure it by the gain from the listing: the runs that fail with the bits only and are solved with the listing. The listing also changes the form in which the bytes are shown, so this gain includes that change, a limit of the design that Section~\ref{sec:threats} states. The second bottleneck is producing a complete, applicable and correct edit once the instructions are given. We read it in the runs that stay unsolved with the listing: most of them fail before the tests, by reaching the output limit or by giving an edit that cannot be applied. The sixth model, Claude Opus 5.5, solves 491 of 500 runs with the bits only, and its few failures sit in single bits of an encoding.

The contributions are the following.
\begin{itemize}
  \item An evaluation design for direct editing of executables: tasks with hidden tests, executables given as bits, five levels of source information, a disassembler listing as a second factor, and a scoring procedure that names the step at which a run fails. The 100 tasks, the 6,000 stored runs and the scripts are available in the artefact that Section~\ref{sec:artefact} describes.
\item A measurement of six models on 100 tasks that separates the effect of the listing, which five models gain, from the effect of source information, which changes the runs solved for two models, in both cases with the bits only.
  \item A characterisation of the decoding bottleneck, with its cost in output tokens, and of the second bottleneck: producing a complete, applicable and correct edit once the instructions are given.
  \item An account of where runs fail, by kind of change and by step, including the byte-order slip, an edit that copies an instruction in the byte order of the listing, the reverse of the file's, and the edits that are shorter than the compiler's.
  \item An account, in Sections~\ref{sec:frontier} and~\ref{sec:hard}, of the failures of the one model that solves nearly all runs with the bits only, and the kinds of task they suggest for a harder evaluation.
  \item Implications for practice, scale and method, which Section~\ref{sec:discussion} states as implications of these results and not as tested claims.
\end{itemize}

\section{Background and related work}
\label{sec:related}

In this section we explain how an executable encodes its instructions, which sets what a direct edit must get right; we review how existing tools change a program without its source, which shows the step that direct editing gives to a language model; and we place our study among prior work on language models below the source code.

\subsection{Executables and their encoding}

An executable holds the machine code of a program together with headers, constants and texts, grouped in sections that the operating system loads into memory. On AArch64, the architecture of this study, every instruction is a 32-bit word, and the file stores it with its least significant byte first.  The word packs several fields, such as the operation, the registers, an immediate value or a condition. A change to one field changes a few bits of the word and leaves the others as they are. The conditional set that compares with sign (\texttt{gt}) and the one that compares without sign (\texttt{hi}), for instance, differ in one bit, a difference that Section~\ref{sec:frontier} meets in the failures of one model.

A disassembler such as \texttt{objdump} reads the sections and prints each instruction as text, with its address and its word in hexadecimal. It prints the word with its most significant byte first, the reverse of the order in the file. Half of the runs of this study receive this output as the disassembler listing, a condition that Section~\ref{sec:prompt} defines.

\subsection{Changing a program without its source}

Binary rewriting changes an executable when its source code is not available, for instrumentation, hardening or patching \citep{wenzl2019binaryrewriting}. Most static rewriters first recover the control flow of the program, because moving an instruction changes the targets of the jumps that reach it. E9Patch avoids that step on x86-64: it leaves the other instructions in place and inserts jumps to new code, called trampolines \citep{duck2020e9patch}. A comparison of ten x86-64 rewriters on 3,344 binaries measures which inputs each tool can rewrite and whether the rewritten programs keep working \citep{schulte2022rewriters}. Scribe patches a binary by recompiling its patched decompiled code \citep{dai2026scribe}. In all of these tools, an algorithm encodes the new instructions. Direct editing gives that step to a language model.

\subsection{Language models below the source code}

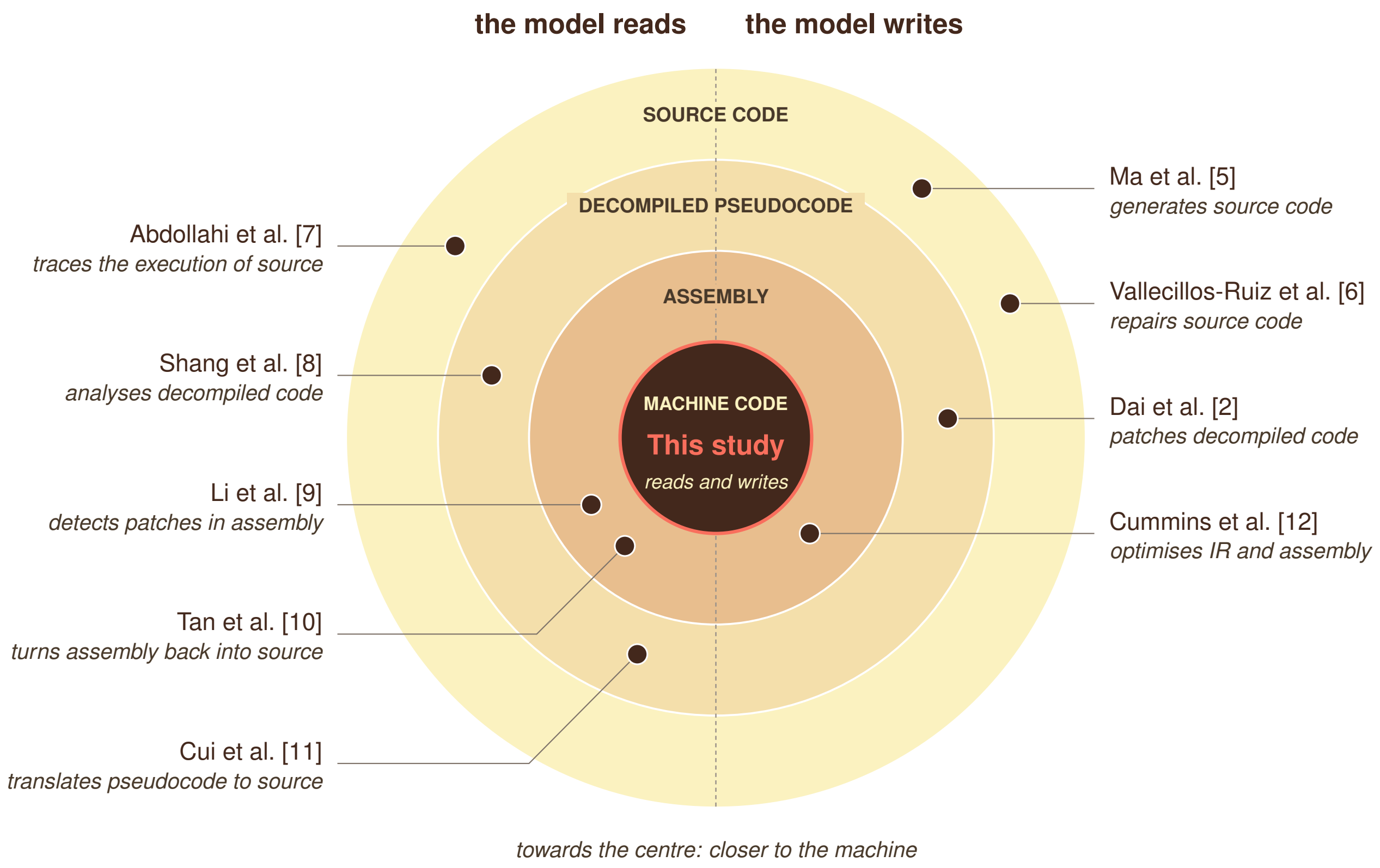
\begin{figure}[t]
  \centering
  \resizebox{\linewidth}{!}{
\begin{tikzpicture}[x=1mm, y=1mm,
  who/.style={inner sep=0, text=coffee, font=\fontfamily{phv}\fontsize{7.2}{8.8}\selectfont},
  what/.style={inner sep=0, text=taupe, font=\fontfamily{phv}\fontsize{6.5}{8}\selectfont\itshape},
  ring/.style={inner xsep=1.2mm, inner ysep=0.5mm, text=taupe, font=\fontfamily{phv}\fontsize{5.8}{7}\selectfont\bfseries},
  head/.style={inner sep=0, text=coffee, font=\fontfamily{phv}\fontsize{8.2}{10}\selectfont\bfseries},
  lead/.style={draw=taupe!70, line width=0.35pt},
  dot/.style={circle, fill=coffee, draw=white, line width=0.5pt, inner sep=0, minimum size=2.1mm}]
\useasboundingbox (0,0) rectangle (139,93);
\definecolor{ringb}{HTML}{F4DFAB}
\definecolor{ringc}{HTML}{E8BE8E}
\fill[lemon] (69.5,46) circle (38.5);
\fill[ringb] (69.5,46) circle (29);
\fill[ringc] (69.5,46) circle (19.5);
\draw[white, line width=0.6pt] (69.5,46) circle (29) (69.5,46) circle (19.5);
\draw[taupe!55, line width=0.4pt, dash pattern=on 1.2pt off 1pt] (69.5,7.5) -- (69.5,84.5);
\node[head, anchor=base east] at (66.5,88.2) {the model reads};
\node[head, anchor=base west] at (72.5,88.2) {the model writes};
\node[ring, fill=lemon] at (69.5,79.75) {SOURCE CODE};
\node[ring, fill=ringb] at (69.5,70.25) {DECOMPILED PSEUDOCODE};
\node[ring, fill=ringc] at (69.5,60.75) {ASSEMBLY};
\fill[coffee] (69.5,46) circle (10);
\draw[coral, line width=1pt] (69.5,46) circle (10);
\node[inner sep=0, text=lemon, font=\fontfamily{phv}\fontsize{5.5}{6.8}\selectfont\bfseries] at (69.5,49.6) {MACHINE CODE};
\node[inner sep=0, text=coral, font=\fontfamily{phv}\fontsize{8.2}{10}\selectfont\bfseries] at (69.5,45.0) {This study};
\node[inner sep=0, text=lemon, font=\fontfamily{phv}\fontsize{5.8}{7}\selectfont\itshape] at (69.5,41.4) {reads and writes};
\draw[lead] (42.3,66) -- (30,66);        \node[dot] at (42.3,66) {};
\node[who, anchor=base east] at (28.5,66.4) {\citet{abdollahi2026reasoning}};
\node[what, anchor=base east] at (28.5,63.4) {traces the execution of source};
\draw[lead] (46.1,52.5) -- (30,52.5);    \node[dot] at (46.1,52.5) {};
\node[who, anchor=base east] at (28.5,52.9) {\citet{shang2025binaryunderstanding}};
\node[what, anchor=base east] at (28.5,49.9) {analyses decompiled code};
\draw[lead] (56.5,39) -- (30,39);        \node[dot] at (56.5,39) {};
\node[who, anchor=base east] at (28.5,39.4) {\citet{li2026binaryspd}};
\node[what, anchor=base east] at (28.5,36.4) {detects patches in assembly};
\draw[lead] (60,34.7) -- (50.8,25.5) -- (30,25.5);  \node[dot] at (60,34.7) {};
\node[who, anchor=base east] at (28.5,25.9) {\citet{tan2024llm4decompile}};
\node[what, anchor=base east] at (28.5,22.9) {turns assembly back into source};
\draw[lead] (61.3,23.4) -- (50,12) -- (30,12);  \node[dot] at (61.3,23.4) {};
\node[who, anchor=base east] at (28.5,12.4) {\citet{cui2026pcodetrans}};
\node[what, anchor=base east] at (28.5,9.4) {translates pseudocode to source};
\draw[lead] (91,72) -- (109,72);         \node[dot] at (91,72) {};
\node[who, anchor=base west] at (110.5,72.4) {\citet{ma2026dlcodegen}};
\node[what, anchor=base west] at (110.5,69.4) {generates source code};
\draw[lead] (100.2,60) -- (109,60);      \node[dot] at (100.2,60) {};
\node[who, anchor=base west] at (110.5,60.4) {\citet{vallecillos2026roundtrip}};
\node[what, anchor=base west] at (110.5,57.4) {repairs source code};
\draw[lead] (93.7,48) -- (109,48);       \node[dot] at (93.7,48) {};
\node[who, anchor=base west] at (110.5,48.4) {\citet{dai2026scribe}};
\node[what, anchor=base west] at (110.5,45.4) {patches decompiled code};
\draw[lead] (79.3,36) -- (109,36);       \node[dot] at (79.3,36) {};
\node[who, anchor=base west] at (110.5,36.4) {\citet{cummins2025llmcompiler}};
\node[what, anchor=base west] at (110.5,33.4) {optimises IR and assembly};
\node[what, anchor=base] at (69.5,2.2) {towards the centre: closer to the machine};
\end{tikzpicture}
  }
  \caption{Where models work on programs. Each ring is a level of program representation, with the machine code at the centre. Each prior work sits on the lowest level its model handles directly: on the left when the model reads at that level, on the right when it writes at it. In the nine works shown, the final bytes come from a compiler or an assembler, or no bytes are written. In this study the model reads and writes the machine code itself.}
  \Description{Three concentric rings around a dark centre. The outer ring is source code, the middle ring is decompiled pseudocode, the inner ring is assembly and the centre is machine code. Nine prior works are marked on the rings, five on the left half where the model reads and four on the right half where the model writes. The centre is labelled This study.}
  \label{fig:map}
\end{figure}

Figure~\ref{fig:map} places prior work by the lowest level of program representation that its model reads or writes. Empirical studies at the level of source code evaluate code generation \citep{ma2026dlcodegen}, program repair \citep{vallecillos2026roundtrip} and the simulation of code execution \citep{abdollahi2026reasoning}.

Below the source code, most work reads. \citet{shang2025binaryunderstanding} evaluate models on binary code understanding, and BinMetric gathers six binary analysis tasks, among them decompilation and the generation of assembly instructions, into one benchmark \citep{shang2025binmetric}. ReSym recovers the names and types of variables and data structures from stripped binaries \citep{xie2024resym}. \citet{li2026binaryspd} detect security patches in binaries, and Nova is a generative language model for assembly code \citep{jiang2025nova}. LLM4Decompile turns assembly back into source code \citep{tan2024llm4decompile}, and PCodeTrans translates decompiled pseudocode into source code \citep{cui2026pcodetrans}.

Some work writes below the source code. LLM Compiler optimises intermediate representation and assembly \citep{cummins2025llmcompiler}, and SuperCoder asks models for assembly programs that run faster than the output of an optimising compiler \citep{wei2026supercoder}. An assembler then turns that text into bytes.

A separate line of work trains models on bytes. bGPT predicts the next byte and is evaluated, among other tasks, on simulating CPU operations and converting music files between formats \citep{wu2024bgpt}. The Large Byte Model extends a language model with a byte vocabulary and answers questions about malware executables from their raw bytes, such as the malware family and the architecture \citep{stortz2026lbm}. Both read bytes, and neither writes an edited executable. None of the works found for this study has a language model edit the bytes of an executable itself, which is the setting this study measures.

\section{Study design}
\label{sec:design}

The study asks three research questions about a model that edits an executable directly, with no compiler.

\begin{itemize}
  \item \textbf{RQ1.} How does the number of runs solved change with the level of source information in the prompt, with and without a disassembler listing?
  \item \textbf{RQ2.} Which kinds of change do models achieve at machine level (change a value, change a decision, insert logic), and for which does the number of runs solved fall?
  \item \textbf{RQ3.} At which step does a failed run fail: locating the place in the sequence, writing the new instruction with the right encoding, or preserving the rest of the program?
\end{itemize}

A run is one model, one task and one condition, scored once. The study has 6,000 runs: 6 models, 100 tasks and 10 conditions per task. Figure~\ref{fig:method} shows the design on one toy task.

\subsection{Tasks}
\label{sec:tasks}

A task is one change request on one small program, with the program's executable and a set of hidden tests. There are 100 tasks in 20 families of 5, and Table~\ref{tab:families} lists the families with one example of each change. A task is named by its family and its number within it: D2\_04 is the fourth task of family D2. The 5 tasks of a family apply the same kind of change with different names, constants and messages.

\begin{table}[t]
\caption{The 20 task families, 5 tasks each. Every task of a family applies the same kind of change with different names and constants. \emph{Edit lands in} says whether the reference solution changes only data bytes or also instruction bytes; \emph{Bytes} is the number of bytes it changes.}
\label{tab:families}
\footnotesize
\setlength{\tabcolsep}{3pt}
\begin{tabular}{llllll}
\toprule
ID & Family & Source before & Source after & Edit lands in & Bytes \\
\midrule
\multicolumn{6}{l}{\textit{Change a value} (6 families, 30 tasks)} \\
V1 & Threshold constant & \texttt{return 56;} & \texttt{return 84;} & instructions & 2 \\
V2 & Multiplier & \texttt{n * 24 + 11} & \texttt{n * 59 + 11} & instructions & 13--25 \\
V3 & Text constant & \texttt{\textquotedbl{}door shut\textquotedbl{}} & \texttt{\textquotedbl{}door ajar\textquotedbl{}} & data & 4--5 \\
V4 & Exit status & \texttt{n \textgreater{} 62 ? 8 : 0} & \texttt{n \textgreater{} 62 ? 26 : 0} & instructions & 2--7 \\
V5 & Table entry & \texttt{\{..., 777, 692\}} & \texttt{\{..., 777, 518\}} & data & 1--2 \\
V6 & Loop count & \texttt{i \textless{} 8} & \texttt{i \textless{} 16} & instructions & 4--29 \\
\addlinespace
\multicolumn{6}{l}{\textit{Change a decision} (7 families, 35 tasks)} \\
D1 & Swapped texts & \texttt{? \textquotedbl{}plain\textquotedbl{} : \textquotedbl{}triple\textquotedbl{}} & \texttt{? \textquotedbl{}triple\textquotedbl{} : \textquotedbl{}plain\textquotedbl{}} & data & 6--12 \\
D2 & Smaller to larger & \texttt{a \textless{} b ? a : b} & \texttt{a \textgreater{} b ? a : b} & instructions & 1 \\
D3 & Inclusive to strict bound & \texttt{n \textgreater{}= 20} & \texttt{n \textgreater{} 20} & instructions & 1 \\
D4 & Reversed test & \texttt{n \textgreater{} 40} & \texttt{n \textless{} 40} & instructions & 2 \\
D5 & Both to either & \texttt{a \textgreater{} 30 \&\& b \textgreater{} 30} & \texttt{a \textgreater{} 30 \textbar{}\textbar{} b \textgreater{} 30} & instructions & 2 \\
D6 & Strict to inclusive bound & \texttt{a \textgreater{} b} & \texttt{a \textgreater{}= b} & instructions & 1 \\
D7 & Negated multiple test & \texttt{n \% 11 == 0} & \texttt{n \% 11 != 0} & instructions & 1 \\
\addlinespace
\multicolumn{6}{l}{\textit{Insert new logic} (7 families, 35 tasks)} \\
I1 & Guard against zero & \texttt{2125 / n} & \texttt{if (n == 0) fail; 2125 / n} & instructions & 49 \\
I2 & Upper clamp & \texttt{r = n * n} & \texttt{r \textgreater{} 4095 ? 4095 : r} & instructions & 17 \\
I3 & Lower clamp at zero & \texttt{r = n - 28} & \texttt{r \textless{} 0 ? 0 : r} & instructions & 10 \\
I4 & Conditional discount & \texttt{r = n * 20} & \texttt{if (n \textgreater{} 9) r -= 119} & instructions & 17--18 \\
I5 & Second range check & \texttt{n \textgreater{} 65} & \texttt{n \textgreater{} 65 \textbar{}\textbar{} n \textless{} 16} & instructions & 16--17 \\
I6 & Absolute difference & \texttt{a - b} & \texttt{a \textgreater{} b ? a - b : b - a} & instructions & 17 \\
I7 & Rounding up & \texttt{n / 6} & \texttt{(n + 5) / 6} & instructions & 5--29 \\
\bottomrule
\end{tabular}
\end{table}

Each family belongs to one of three kinds of change. In a \emph{value} change (6 families, 30 tasks) the program must use a different constant, such as a threshold of 84 instead of 56. In a \emph{decision} change (7 families, 35 tasks) a test or a choice must change, such as a smaller-than test that becomes a greater-than test. In an \emph{insert logic} change (7 families, 35 tasks) the program must gain behaviour it lacks, such as a guard against division by zero or an upper clamp. The kind of change is the factor of RQ2.

Each task also has a reference solution: the bytes that differ between the compiled original and the compiled modified source. It overwrites bytes and never changes the length of the file. It changes between 1 and 49 bytes, a count that the last column of Table~\ref{tab:families} gives for each family. For 15 tasks (families V3, V5 and D1) the reference solution changes only data bytes, such as texts and table entries. For the other 85 it also changes instruction bytes.

The tasks come from fixed generation rules with a fixed random seed.\footnote{See \texttt{gen\_tasks.py} in the artefact that Section~\ref{sec:artefact} describes.} For each family the generator writes 8 candidates with random parameters, compiles them, and keeps the first 5 that pass the validity checks. A candidate is valid when its reference solution has the same file length as the original, the original executable fails at least one test of the new behaviour, passes every test of the old behaviour, the old-behaviour tests exist, the reference solution passes every test, and the outputs that the change request states hold on the programs.
\subsection{Executables}
\label{sec:build}

Each task has two C programs of 10 to 15 lines, the original and the modified one. Both call a runtime of 28 lines (\texttt{rt.c}) that holds the start-up code, a system call to write, and three small helper functions. The executables are built for AArch64 Linux with GCC 12.2.0 in a container, with the flags \texttt{-O1 -static -nostdlib -ffreestanding -fno-inline -falign-functions=64}, among others, and then stripped of symbol names. The last flag starts every function at a multiple of 64 bytes, so a function is followed by padding made of no-operation instructions up to the next multiple of 64 bytes. Without a C library the executables carry no library code.

\begin{figure}[t]
  \centering
  \includegraphics[width=\linewidth]{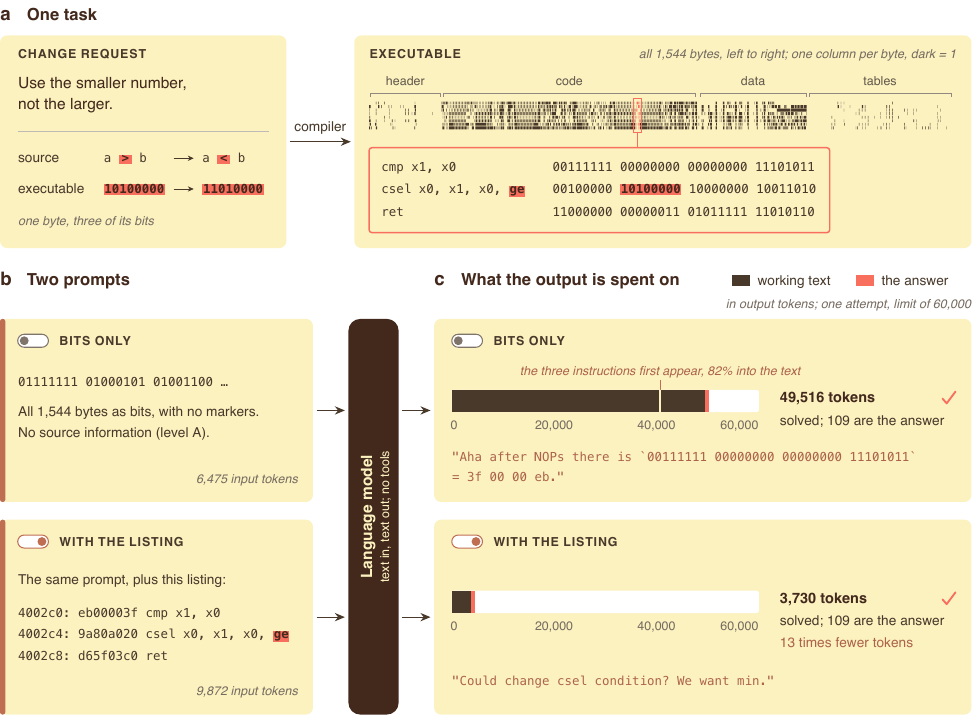}
  \caption{The decoding bottleneck in one task of the study, D2\_04 at level A. (a) The requested change and the three instructions that decide it; the fix is one byte. (b) With the bits only, the three instructions sit unmarked among 1,544 bytes; with the listing, each one is named. (c) Output tokens of the same model, DeepSeek-V4.1-Flash, in one attempt: with the bits only the task is solved after 49,516 tokens, and the three instructions first appear 82\% into the text; with the listing it is solved after 3,730. The quoted lines are verbatim from the stored output.}
  \Description{Three panels. Panel a shows the change request and the executable as a tape of bytes with three instructions enlarged. Panel b shows a prompt with the bits only and a prompt with the disassembler listing. Panel c shows two bars of output tokens: solved at 49,516 with the bits only and solved at 3,730 with the listing.}
  \label{fig:bottleneck}
\end{figure}

The 100 executables are between 1,480 and 1,592 bytes, with a mean of 1,526. The size is a design choice. It is small enough that a prompt holds the whole file as bits and large enough that the instruction to change sits among bytes that are not instructions, such as headers and constants. Figure~\ref{fig:bottleneck} shows what this means for one task. The requested change depends on three instructions, and the fix is one byte (panel a). With the bits only, nothing in the prompt marks those instructions among the rest of the file; with the listing, each one is named (panel b). The same model solves the task in both cases, but with the bits only it produces far more output before it answers (panel c). That difference is what the decoding bottleneck costs in this task.

\subsection{Prompt and conditions}
\label{sec:prompt}

The executable reaches the model in a prompt, and the conditions of the study are variations of that prompt. Each run sends one user message, with no system message and no tools. The message holds, in this order: a short description of the format (AArch64 Linux, little endian, statically linked, no C library, no symbol names, and the statement that the result of the edits is executed as is); the change request in one or two sentences; the source information, if the level has any; the disassembler listing, if the condition has it; the executable as bits; and the answer format. The prompt of a run is the same text for every model.

The executable is given as bits: 8 binary digits per byte, in file order, most significant digit first, with one space between bytes. Bits are the lowest-level form of the file: without the listing, every instruction must be located as a pattern of binary digits. A file of 1,526 bytes takes about 13,700 characters (9 per byte).

The \emph{source information} has five levels, which panel (b) of Figure~\ref{fig:method} labels A to E:
\begin{itemize}
  \item \textbf{A}, no source information.
  \item \textbf{B}, the header files: the declarations of the runtime and of the program, with no function bodies. A one-line comment beside each declaration says what the function returns; in 22 of the 35 insert logic tasks it states the computation with its constant (for example, \emph{2125 divided by n}).
  \item \textbf{C}, an older version of the source file. It is not the version that produced the executable, and in 65 of the 100 tasks it differs from \texttt{main.c} on a line that the requested change also touches.
  \item \textbf{D}, the source file \texttt{main.c} as compiled into the executable.
  \item \textbf{E}, the desired change as a diff against \texttt{main.c}. The rest of the source is not given.
\end{itemize}

The \emph{disassembler listing} is the output of \texttt{objdump} on the executable, pasted as text: the section table, the disassembly of the code and a dump of the read-only data. The listing has no symbol names, because the executable has none, and the task name is replaced by \texttt{prog}. It prints each 4-byte instruction as one hexadecimal number, most significant byte first, while the file stores the same instruction least significant byte first. This difference matters for the byte-order slips of Section~\ref{sec:further}.

Five levels and two states of the listing give 10 conditions per task: five with the bits only, one for each level, and five with the listing. Levels serve RQ1 as one factor. The listing serves RQ1 as the second factor, and serves RQ3 because it adds the instructions, their addresses and the sections as readable text without changing the executable.

\subsection{Answer format and application of edits}
\label{sec:answer}

The model answers with search-and-replace edits on the bits. The answer ends with one JSON object, \texttt{\{"edits": [\ldots]\}}, whose list holds pairs of \texttt{old} and \texttt{new}. Each \texttt{old} is a run of consecutive bytes copied from the executable, written as 8-bit groups. It must occur exactly once in the file, and \texttt{new} must have the same number of bytes. Edits must not overlap. The parser takes the last JSON object with an \texttt{edits} list in the final answer; text produced before the answer is ignored. Pairs of hexadecimal digits are accepted in place of 8-bit groups, because the listing is in hexadecimal and rejecting them would score the format instead of the edit.

All edits are located in the original executable and applied together. An edit that is not made of 8-bit groups or hexadecimal pairs, that has unequal lengths, whose \texttt{old} does not occur or occurs more than once, or that overlaps another, makes the answer not applicable.

\subsection{Hidden tests and scoring}
\label{sec:scoring}

An answer that can be applied gives an edited executable, and the hidden tests decide whether it does what was asked. The tests are kept out of the prompt. Each task has 5 to 8 tests, 6.2 on average: 1 to 5 tests of the new behaviour and 1 to 5 tests of the old behaviour that must not change. A test runs the edited executable with some arguments in a container with no network, a limit of 2 seconds, and compares the output and the exit status with the expected ones. The expected values come from running the compiled modified source.

\begin{figure}[t]
  \centering
  \resizebox{\linewidth}{!}{
\begin{tikzpicture}[x=1mm, y=-1mm,
  q/.style={anchor=base west, inner sep=0, text=coffee, font=\fontfamily{phv}\fontsize{8.8}{10.5}\selectfont\bfseries},
  d/.style={anchor=base west, inner sep=0, text=taupe, font=\fontfamily{phv}\fontsize{6.6}{8.2}\selectfont},
  nm/.style={anchor=base west, inner sep=0, text=coffee, font=\fontfamily{phv}\fontsize{8.2}{10}\selectfont\bfseries},
  ds/.style={anchor=base west, inner sep=0, text=taupe, font=\fontfamily{phv}\fontsize{6.8}{8.2}\selectfont\itshape},
  yn/.style={inner sep=0, text=taupe, font=\fontfamily{phv}\fontsize{6.5}{8}\selectfont\itshape},
  stepname/.style={anchor=base east, inner sep=0, text=cinnamon, font=\fontfamily{phv}\fontsize{6}{7.5}\selectfont\bfseries},
  num/.style={circle, fill=coffee, inner sep=0, minimum size=4.2mm, text=lemon, font=\fontfamily{phv}\fontsize{8}{8.8}\selectfont\bfseries},
  flow/.style={draw=taupe, line width=0.6pt, -{Stealth[length=1.7mm, width=1.4mm]}},
  chip/.style={draw=taupe!75, line width=0.45pt, rounded corners=1.2mm, fill=white}]
\useasboundingbox (0,0) rectangle (139,102);
\foreach \k/\y/\question/\detail in {
  1/4/{Did the output finish?}/{the output ended before the limit of 60,000 tokens},
  2/24/{Is there an edit?}/{the answer holds a list of old and new bits},
  3/44/{Can the edit be applied?}/{every old run occurs once in the file; lengths match},
  4/64/{Do all hidden tests pass?}/{the edited file is run on tests kept out of the prompt}} {
  \fill[lemon, rounded corners=1.2mm] (0,\y) rectangle (66,\y+15);
  \node[num] at (5.6,\y+5.3) {\k};
  \node[q] at (10.6,\y+6.6) {\question};
  \node[d] at (10.6,\y+11.4) {\detail};
}
\foreach \y in {19, 39, 59} {
  \draw[flow] (33,\y+0.4) -- (33,\y+4.6);
  \node[yn, anchor=base west] at (34.6,\y+3.4) {yes};
}
\draw[flow] (33,79.4) -- (33,85.6);
\node[yn, anchor=base west] at (34.6,83.4) {yes};
\fill[coral, rounded corners=1.2mm] (10,86) rectangle (56,97);
\node[anchor=base, inner sep=0, text=coffee, font=\fontfamily{phv}\fontsize{9.2}{11.2}\selectfont\bfseries] at (33,90.9) {Solved};
\node[anchor=base, inner sep=0, text=coffee, font=\fontfamily{phv}\fontsize{6.8}{8.2}\selectfont] at (33,94.6) {every hidden test passes, new and old};
\foreach \y/\name/\desc in {
  5.5/{Cut}/{the limit came before any answer},
  25.5/{No edit}/{the answer gives no edit to apply},
  45.5/{Not applicable}/{the file cannot be changed as asked}} {
  \draw[flow] (66.5,\y+6) -- (75.5,\y+6);
  \node[yn, anchor=base] at (71,\y+4.6) {no};
  \draw[chip] (76,\y) rectangle (139,\y+12);
  \node[nm] at (79,\y+5.2) {\name};
  \node[ds] at (79,\y+9.4) {\desc};
}
\draw[flow] (66.5,71.5) -- (71.6,71.5);
\node[yn, anchor=base] at (69,70.1) {no};
\draw[taupe, line width=0.6pt] (72,66) -- (72,93);
\foreach \y/\name/\desc/\stepn in {
  60/{Wrong place}/{the edit misses the code that had to change}/{LOCATING},
  73.5/{Wrong change}/{right place, but the new behaviour fails}/{WRITING},
  87/{Broke the rest}/{the new behaviour works, the old one fails}/{PRESERVING}} {
  \draw[taupe, line width=0.6pt] (72,\y+6) -- (76,\y+6);
  \draw[chip] (76,\y) rectangle (139,\y+12);
  \fill[cinnamon, rounded corners=0.4mm] (76,\y+0.2) rectangle (76.9,\y+11.8);
  \node[nm] at (79.6,\y+5.2) {\name};
  \node[ds] at (79.6,\y+9.4) {\desc};
  \node[stepname] at (136.6,\y+5.0) {\stepn};
}
\end{tikzpicture}
  }
  \caption{How each run is scored. Four checks are applied in order, and the first one that fails names the outcome. A run is solved only when every hidden test passes. When a test fails, the outcome records the step that failed: locating the code to change, writing the new instruction, or preserving the rest of the program.}
  \Description{A column of four checks: did the output finish, is there an edit, can the edit be applied, do all hidden tests pass. A failed check leads to an outcome on the right: cut, no edit, not applicable. A failed test leads to one of three outcomes: wrong place, wrong change, or broke the rest. Passing all four checks leads to solved.}
  \label{fig:outcomes}
\end{figure}
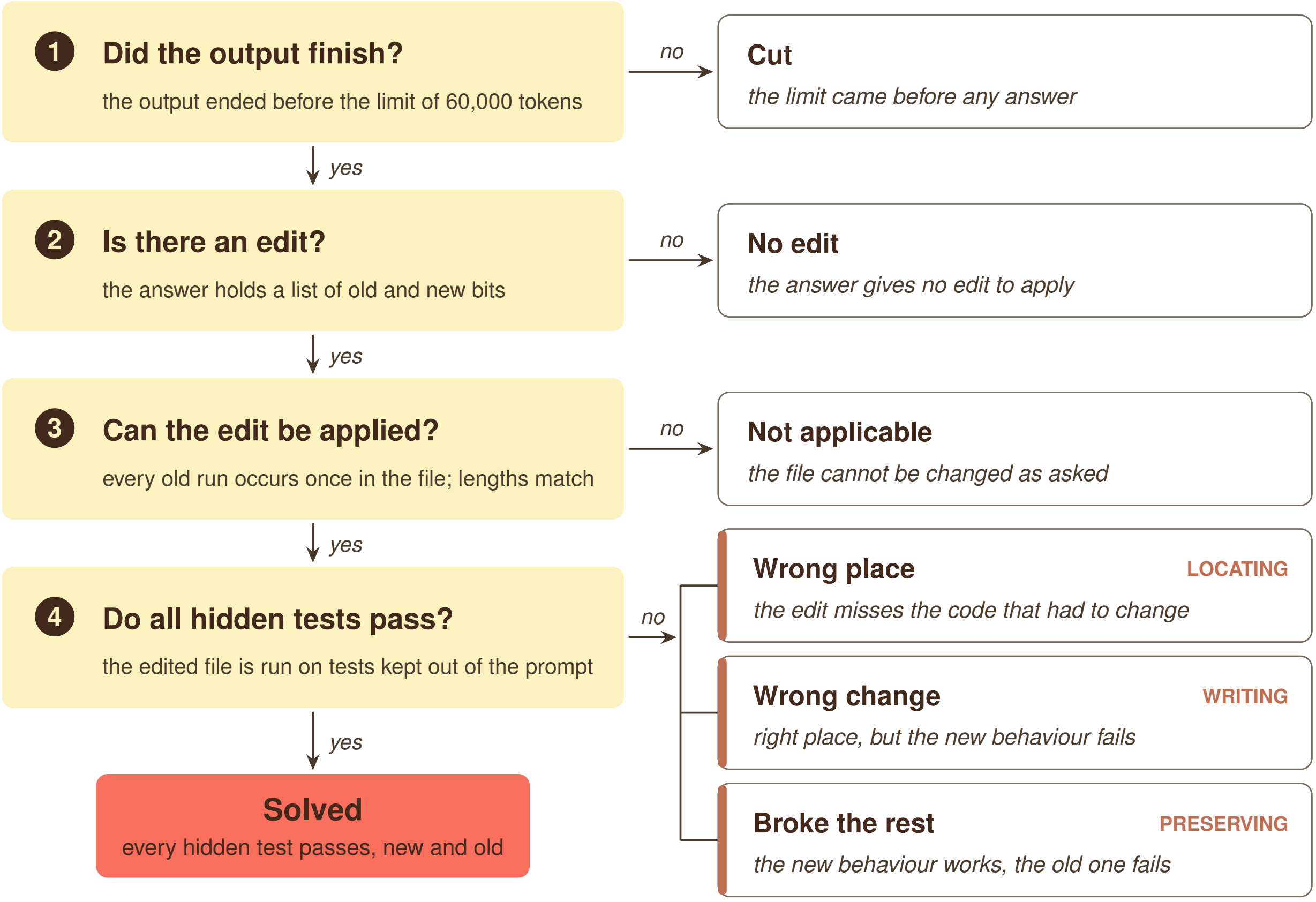

Figure~\ref{fig:outcomes} shows how each run gets its outcome. A run goes through four checks in order (the column on the left), and the first check that fails leads to an outcome on the right. A run that passes all four is \emph{solved}: every hidden test passes. The first check asks whether the output finished before the output limit; if not, the outcome is \emph{cut}. The second asks whether the answer holds an edit; if not, \emph{no edit}. The third asks whether the edit can be applied; if not, \emph{not applicable}. The fourth runs the tests.

A failed test is attributed to one of three steps. The region of a task is the function, or the data, that the reference solution changes (the symbol that holds a changed byte, or 8 bytes on each side when no symbol does). If the edit changes no byte of the region, the outcome is \emph{wrong place} and the step is locating. If it changes the region but a test of the new behaviour fails, it is \emph{wrong change} and the step is writing. If the new behaviour works but a test of the old behaviour fails, it is \emph{broke the rest} and the step is preserving. These three outcomes answer RQ3.

\subsection{Models and configuration}
\label{sec:models}

The tasks, the prompt and the scoring are the same for every model. We called six models from six developers through two providers. Nebius served gpt-oss-120b, Qwen3-235B, Nemotron-3 Super, MiniMax-M3 and DeepSeek-V4.1-Flash, and Anthropic served Claude Opus 5.5. Table~\ref{tab:models} names the developer of each model next to its headline results, and Section~S1 of the supplementary material completes the setup with identifiers, prices, tokens and cost. Sampling parameters are the provider's defaults, and Claude Opus 5.5 runs at its default effort level.

Each run is one attempt, with no feedback and no revision. The output limit is 60,000 tokens for all models, although the provider stops Nemotron-3 Super at 32,768. Because every run has one attempt, its stored output also tells what a smaller limit would give: the run counts as solved at that limit when its complete answer fits within it. Section~\ref{sec:further} uses this to show how the runs solved grow with the limit.

\subsection{Statistical analysis}
\label{sec:stats}

We answer RQ1 with two paired tests, and Table~\ref{tab:tests} reports both for every model. The first test concerns the listing. Each of the 500 pairs of runs of a model has the same task and level of source information and differs only in the listing. We compare the pairs solved only with the listing with those solved only with the bits, using the test of \citet{mcnemar1947} in its exact two-sided form. The second test concerns source information. The $Q$ test of \citet{cochran1950} compares the five levels with the 100 tasks as blocks, for each model and each state of the listing, with 4 degrees of freedom. This gives 18 tests, three for each of the six models, and we adjust them with the method of \citet{holm1979}.

The five pairs of a task share the task, so the McNemar test treats as independent pairs that are not. The results therefore also give the number of tasks that gain and lose with the listing. All other results are counts of runs. For the reader who wants to look further, the supplementary material adds paired differences between levels with intervals, the listing at a common output limit, and the two comparisons of RQ2 with the family as the unit.

\subsection{Scope and artefact}
\label{sec:artefact}

The design sets the scope of the results, and the artefact lets a reader check them. The executables are small by design: a prompt grows in proportion to the file, at 9 characters per byte, and the results apply to files of this size.

Generating the tasks gives the measurements four properties. Each task has a reference solution and hidden tests, so a solved run has a precise meaning. The location of the change is known, so a failed test can be attributed to the wrong place or to a wrong change. The tasks were not published before the runs, so they cannot be in the training data of any model. The whole executable fits in the prompt as bits. The tasks are also small as changes: each one changes one line of the source, where \citet{li2017securitypatches} report a median of 7 changed lines for a security fix and 16 for other bug fixes. We do not claim that the tasks are representative of real fixes.

The artefact holds the 100 tasks with their sources and hidden tests, the executables and listings, the 6,000 stored runs, the 250 control runs of Section~\ref{sec:frontier}, and the scripts that generate the tasks, score a run, and write every table and figure of Section~\ref{sec:results}. Each stored run carries a fingerprint of its input: the model, task, condition, output limit, and the SHA-256 hashes of the prompt and of the executable. A stored result is reused only if the fingerprint matches. Rerunning the models costs about as much as Supplementary Table~S1 shows, and the responses can differ between calls. The data availability statement and Supplementary Figure~S3 describe the archive.



\section{Results}
\label{sec:results}

A language model can change a small compiled program by editing its bits, and how often it succeeds depends on the model and on the disassembler listing. Table~\ref{tab:models} shows how wide the range is, with the runs solved and the output tokens of each model. Claude Opus 5.5 solves 491 of its 500 runs with the bits only. DeepSeek-V4.1-Flash solves 238 with the bits only and 490 with the listing. The other four, gpt-oss-120b, Qwen3-235B, Nemotron-3 Super and MiniMax-M3, gain with the listing too, yet they fail most of their runs in both conditions; below, \emph{the four models} means these four.

We take the three research questions in turn. Section~\ref{sec:rq1} measures what the listing and the source information change, Section~\ref{sec:rq2} asks which kinds of change are solved, and Section~\ref{sec:rq3} finds the step at which the other runs fail. Section~\ref{sec:further} then looks at the form of the edits and at the output limit, and Section~\ref{sec:frontier} examines the few failures of Claude Opus 5.5.

\begin{table}[t]
\caption{Models and headline results on 100 tasks. Each model ran 500 runs with the bits only and 500 with the disassembler listing, one attempt per run and up to 60,000 output tokens (32,768 for Nemotron-3 Super, the limit of its provider). \emph{Cut} counts the runs with the bits only that reached the output limit before giving an answer.}
\label{tab:models}
\footnotesize
\setlength{\tabcolsep}{4pt}
\begin{tabular}{llrrrrr}
\toprule
 & & \multicolumn{2}{c}{Solved runs} & & \multicolumn{2}{c}{Mean output tokens} \\
\cmidrule(lr){3-4}\cmidrule(lr){6-7}
Model & Developer & bits only & with listing & Cut & bits only & with listing \\
\midrule
gpt-oss-120b & OpenAI & 89 (18\%) & 190 (38\%) & 0 & 3,260 & 6,500 \\
Qwen3-235B & Alibaba & 69 (14\%) & 112 (22\%) & 64 & 20,171 & 12,952 \\
Nemotron-3 Super & NVIDIA & 25 (5\%) & 90 (18\%) & 450 & 30,104 & 26,761 \\
MiniMax-M3 & MiniMax & 47 (9\%) & 142 (28\%) & 428 & 54,792 & 42,575 \\
DeepSeek-V4.1-Flash & DeepSeek & 238 (48\%) & 490 (98\%) & 254 & 42,985 & 7,240 \\
Claude Opus 5.5 & Anthropic & 491 (98\%) & 498 (100\%) & 0 & 1,363 & 1,058 \\
\bottomrule
\end{tabular}
\end{table}

\subsection{RQ1: Source information and the disassembler listing}
\label{sec:rq1}

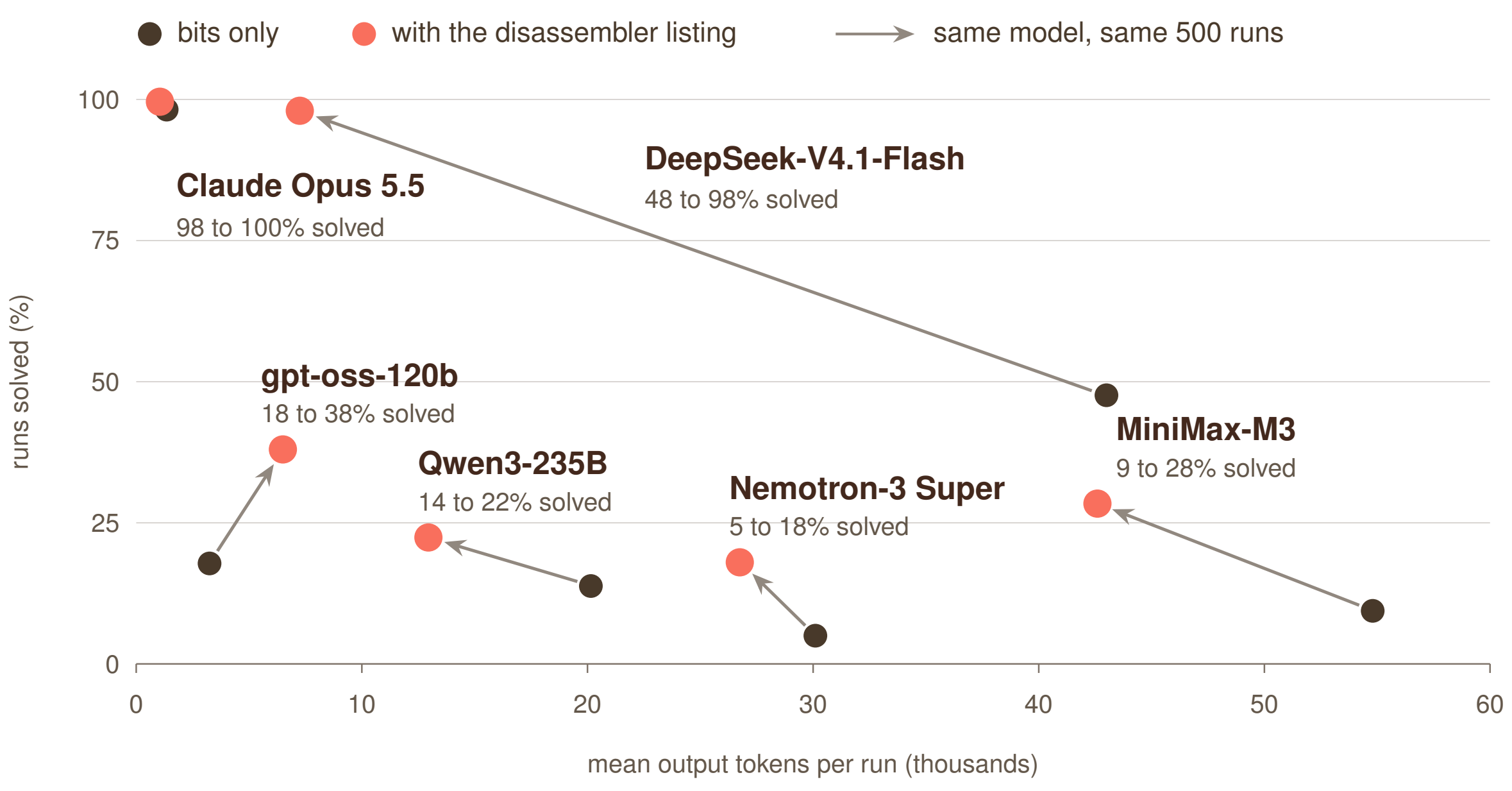
\begin{figure}[t]
  \centering
  \resizebox{\linewidth}{!}{
\begin{tikzpicture}[x=1mm, y=-1mm]
\useasboundingbox (0,0) rectangle (139,71);
\draw[rgrid] (13,59) -- (133,59);
\node[rtick, anchor=east] at (11.6,59) {0};
\draw[rgrid] (13,46.5) -- (133,46.5);
\node[rtick, anchor=east] at (11.6,46.5) {25};
\draw[rgrid] (13,34) -- (133,34);
\node[rtick, anchor=east] at (11.6,34) {50};
\draw[rgrid] (13,21.5) -- (133,21.5);
\node[rtick, anchor=east] at (11.6,21.5) {75};
\draw[rgrid] (13,9) -- (133,9);
\node[rtick, anchor=east] at (11.6,9) {100};
\draw[raxis] (13,59) -- (133,59);
\draw[raxis] (13,59) -- (13,59.8);
\node[rtick, anchor=base] at (13,63.3) {0};
\draw[raxis] (33,59) -- (33,59.8);
\node[rtick, anchor=base] at (33,63.3) {10};
\draw[raxis] (53,59) -- (53,59.8);
\node[rtick, anchor=base] at (53,63.3) {20};
\draw[raxis] (73,59) -- (73,59.8);
\node[rtick, anchor=base] at (73,63.3) {30};
\draw[raxis] (93,59) -- (93,59.8);
\node[rtick, anchor=base] at (93,63.3) {40};
\draw[raxis] (113,59) -- (113,59.8);
\node[rtick, anchor=base] at (113,63.3) {50};
\draw[raxis] (133,59) -- (133,59.8);
\node[rtick, anchor=base] at (133,63.3) {60};
\node[rtick, anchor=base, rotate=90] at (3.4,34) {runs solved (\%)};
\node[rtick, anchor=base] at (73,68.6) {mean output tokens per run (thousands)};
\fill[taupe] (14.2,3.2) circle (1.05); \node[rlab, anchor=base west] at (16.6,3.9) {bits only};
\fill[coral] (33.2,3.2) circle (1.05); \node[rlab, anchor=base west] at (35.6,3.9) {with the disassembler listing};
\draw[taupe!60, line width=0.7pt, -{Stealth[length=1.9mm, width=1.5mm]}] (75,3.2) -- (82,3.2); \node[rlab, anchor=base west] at (83.6,3.9) {same model, same 500 runs};
\draw[taupe!60, line width=0.8pt, -{Stealth[length=1.9mm, width=1.5mm]}] (20.2,49) -- (25.2,41.3);
\fill[taupe] (19.5,50.1) circle (1.05);
\fill[coral] (26,40) circle (1.25);
\node[rttl, anchor=base west] at (24,34.4) {gpt-oss-120b};
\node[rtick, anchor=base west] at (24,37.6) {18 to 38\% solved};
\draw[taupe!60, line width=0.8pt, -{Stealth[length=1.9mm, width=1.5mm]}] (52.1,51.7) -- (40.3,48.2);
\fill[taupe] (53.3,52.1) circle (1.05);
\fill[coral] (38.9,47.8) circle (1.25);
\node[rttl, anchor=base west] at (37.9,42.2) {Qwen3-235B};
\node[rtick, anchor=base west] at (37.9,45.4) {14 to 22\% solved};
\draw[taupe!60, line width=0.8pt, -{Stealth[length=1.9mm, width=1.5mm]}] (72.3,55.6) -- (67.6,51);
\fill[taupe] (73.2,56.5) circle (1.05);
\fill[coral] (66.5,50) circle (1.25);
\node[rttl, anchor=base west] at (65.5,44.4) {Nemotron-3 Super};
\node[rtick, anchor=base west] at (65.5,47.6) {5 to 18\% solved};
\draw[taupe!60, line width=0.8pt, -{Stealth[length=1.9mm, width=1.5mm]}] (121.4,53.8) -- (99.5,45.3);
\fill[taupe] (122.6,54.3) circle (1.05);
\fill[coral] (98.2,44.8) circle (1.25);
\node[rttl, anchor=base west] at (99.8,39.2) {MiniMax-M3};
\node[rtick, anchor=base west] at (99.8,42.4) {9 to 28\% solved};
\draw[taupe!60, line width=0.8pt, -{Stealth[length=1.9mm, width=1.5mm]}] (97.7,34.8) -- (28.9,10.5);
\fill[taupe] (99,35.2) circle (1.05);
\fill[coral] (27.5,10) circle (1.25);
\node[rttl, anchor=base west] at (58,15.2) {DeepSeek-V4.1-Flash};
\node[rtick, anchor=base west] at (58,18.6) {48 to 98\% solved};
\draw[taupe!60, line width=0.8pt, -{Stealth[length=1.9mm, width=1.5mm]}] (14.9,8.9) -- (16.1,10.3);
\fill[taupe] (15.7,9.9) circle (1.05);
\fill[coral] (15.1,9.2) circle (1.25);
\node[rttl, anchor=base west] at (16.5,17.6) {Claude Opus 5.5};
\node[rtick, anchor=base west] at (16.5,21.1) {98 to 100\% solved};
\end{tikzpicture}
  }
  \caption{What the disassembler listing changes, for each model. Each arrow joins the two conditions of one model: the 500 runs with the bits only and the same runs with the listing. The vertical move is the change in runs solved and the horizontal move is the change in the length of the output.}
  \Description{A plane with output tokens on the horizontal axis and runs solved on the vertical axis. Six arrows, one per model, each from the point with the bits only to the point with the disassembler listing.}
  \label{fig:headline}
\end{figure}

\paragraph{The listing raises the runs solved for five of the six models}
Figure~\ref{fig:headline} and Table~\ref{tab:models} show the change for each model. DeepSeek-V4.1-Flash goes from 48 to 98 percent of its runs solved, a gain of 50 points. The other four models gain 8 to 20 points and fail most of their runs with the listing. Claude Opus 5.5 solves 98 percent of its runs with the bits only and nearly all with the listing.

\paragraph{The gain is paired and holds at task level}
Each pair of runs in Table~\ref{tab:tests} has the same task and level of source information and differs only in the listing. For five models, the pairs solved only with the listing far outnumber those solved only with the bits, and their five adjusted $p$-values are below 0.001. Claude Opus 5.5 has 9 pairs solved only with the listing and 2 only with the bits (unadjusted $p = 0.065$, adjusted $p = 0.720$). At task level the direction is the same. The columns \emph{tasks up} and \emph{tasks down} of the table count the tasks on which the listing raises or lowers the number of levels solved: for each of the five models, it raises it on many more tasks than it lowers.

\begin{table}[t]
\caption{Paired tests on the 100 tasks. \emph{Listing}: of the 500 pairs of runs that differ only in the disassembler listing, those solved only with the listing and only with the bits, with an exact McNemar test; \emph{tasks up} and \emph{tasks down} count the tasks on which the listing raises or lowers the number of levels solved. \emph{Source information}: Cochran's $Q$ over the five levels (4 degrees of freedom), with the 100 tasks as blocks. All $p$-values are Holm-adjusted over the 18 tests of the table.}
\label{tab:tests}
\footnotesize
\setlength{\tabcolsep}{4pt}
\begin{tabular}{lrrrrrrrrr}
\toprule
 & \multicolumn{5}{c}{Listing} & \multicolumn{4}{c}{Source information} \\
\cmidrule(lr){2-6}\cmidrule(lr){7-10}
 & only with & only with & & tasks & tasks & \multicolumn{2}{c}{bits only} & \multicolumn{2}{c}{with listing} \\
\cmidrule(lr){7-8}\cmidrule(lr){9-10}
Model & listing & bits & $p$ & up & down & $Q$ & $p$ & $Q$ & $p$ \\
\midrule
gpt-oss-120b & 109 & 8 & $<$0.001 & 39 & 3 & 4.9 & 1.000 & 2.1 & 1.000 \\
Qwen3-235B & 57 & 14 & $<$0.001 & 29 & 4 & 4.5 & 1.000 & 4.2 & 1.000 \\
Nemotron-3 Super & 74 & 9 & $<$0.001 & 31 & 2 & 2.2 & 1.000 & 7.2 & 1.000 \\
MiniMax-M3 & 109 & 14 & $<$0.001 & 56 & 6 & 1.7 & 1.000 & 1.5 & 1.000 \\
DeepSeek-V4.1-Flash & 256 & 4 & $<$0.001 & 85 & 1 & 27.9 & $<$0.001 & 5.0 & 1.000 \\
Claude Opus 5.5 & 9 & 2 & 0.720 & 6 & 1 & 19.2 & 0.009 & 4.0 & 1.000 \\
\bottomrule
\end{tabular}
\end{table}

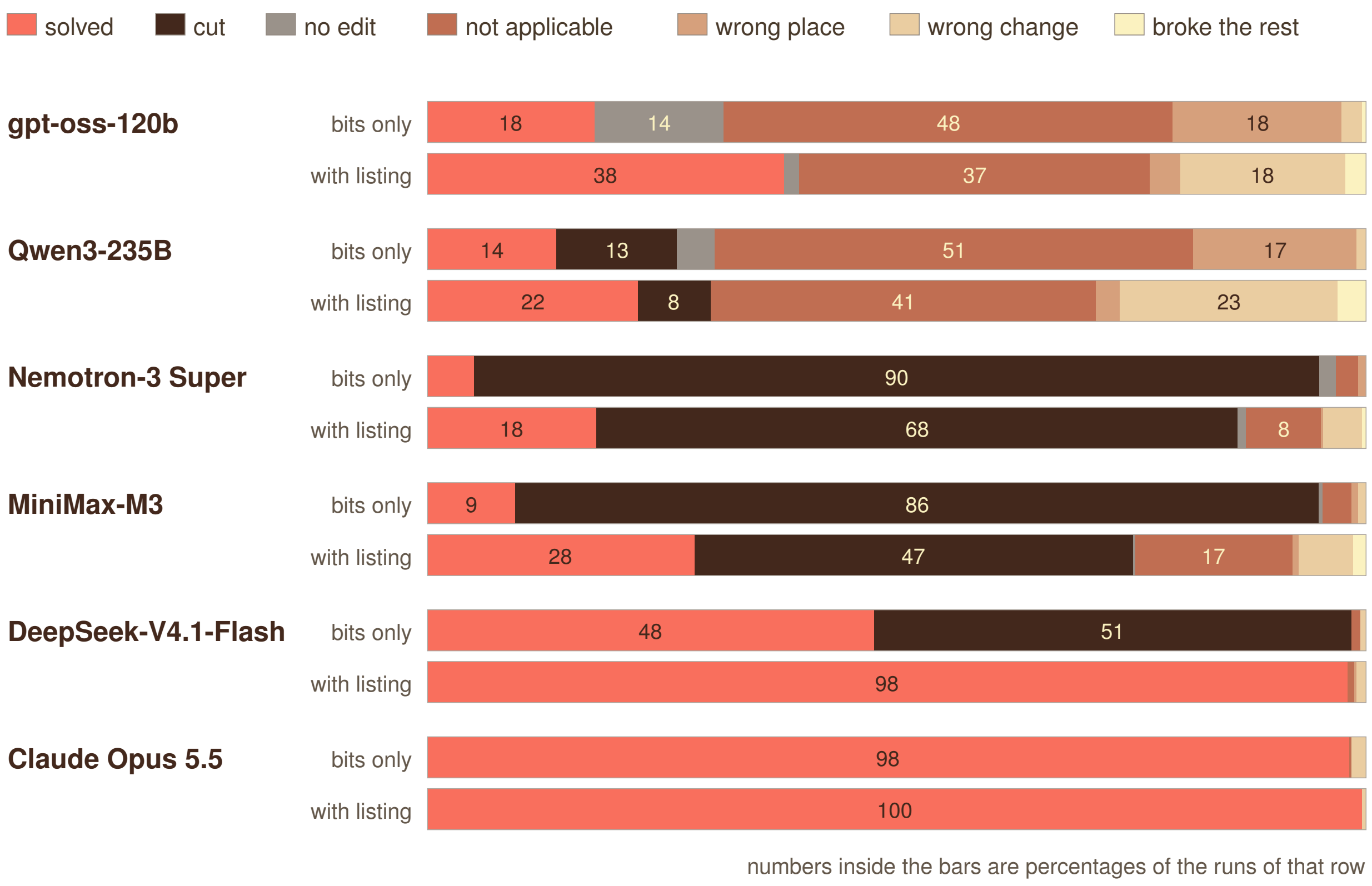
\begin{figure}[t]
  \centering
  \resizebox{\linewidth}{!}{
\begin{tikzpicture}[x=1mm, y=-1mm]
\useasboundingbox (0,0) rectangle (139,94);
\draw[taupe!60, line width=0.25pt, fill=coral] (0,1.0) rectangle (3,3.2);
\node[rlab, anchor=base west] at (3.8,3.2) {solved};
\draw[taupe!60, line width=0.25pt, fill=coffee] (15.2,1.0) rectangle (18.2,3.2);
\node[rlab, anchor=base west] at (19,3.2) {cut};
\draw[taupe!60, line width=0.25pt, fill=taupe!55] (26.5,1.0) rectangle (29.5,3.2);
\node[rlab, anchor=base west] at (30.3,3.2) {no edit};
\draw[taupe!60, line width=0.25pt, fill=cinnamon] (43,1.0) rectangle (46,3.2);
\node[rlab, anchor=base west] at (46.8,3.2) {not applicable};
\draw[taupe!60, line width=0.25pt, fill=cinnamon!62!lemon] (68.6,1.0) rectangle (71.6,3.2);
\node[rlab, anchor=base west] at (72.4,3.2) {wrong place};
\draw[taupe!60, line width=0.25pt, fill=cinnamon!28!lemon] (90.3,1.0) rectangle (93.3,3.2);
\node[rlab, anchor=base west] at (94.1,3.2) {wrong change};
\draw[taupe!60, line width=0.25pt, fill=lemon] (113.3,1.0) rectangle (116.3,3.2);
\node[rlab, anchor=base west] at (117.1,3.2) {broke the rest};
\node[rttl, anchor=base west] at (0,13.2) {gpt-oss-120b};
\node[rtick, anchor=base east] at (41.5,13.1) {bits only};
\fill[coral] (43,10) rectangle (60.1,14.2);
\node[rval, text=coffee, anchor=base] at (51.5,13) {18};
\fill[taupe!55] (60.1,10) rectangle (73.3,14.2);
\node[rval, text=lemon, anchor=base] at (66.7,13) {14};
\fill[cinnamon] (73.3,10) rectangle (119.2,14.2);
\node[rval, text=lemon, anchor=base] at (96.3,13) {48};
\fill[cinnamon!62!lemon] (119.2,10) rectangle (136.5,14.2);
\node[rval, text=coffee, anchor=base] at (127.9,13) {18};
\fill[cinnamon!28!lemon] (136.5,10) rectangle (138.6,14.2);
\fill[lemon] (138.6,10) rectangle (139,14.2);
\draw[taupe!45, line width=0.25pt] (43,10) rectangle (139,14.2);
\node[rtick, anchor=base east] at (41.5,18.4) {with listing};
\fill[coral] (43,15.3) rectangle (79.5,19.5);
\node[rval, text=coffee, anchor=base] at (61.2,18.3) {38};
\fill[taupe!55] (79.5,15.3) rectangle (81,19.5);
\fill[cinnamon] (81,15.3) rectangle (116.9,19.5);
\node[rval, text=lemon, anchor=base] at (99,18.3) {37};
\fill[cinnamon!62!lemon] (116.9,15.3) rectangle (120,19.5);
\fill[cinnamon!28!lemon] (120,15.3) rectangle (136.9,19.5);
\node[rval, text=coffee, anchor=base] at (128.4,18.3) {18};
\fill[lemon] (136.9,15.3) rectangle (139,19.5);
\draw[taupe!45, line width=0.25pt] (43,15.3) rectangle (139,19.5);
\node[rttl, anchor=base west] at (0,26.2) {Qwen3-235B};
\node[rtick, anchor=base east] at (41.5,26.1) {bits only};
\fill[coral] (43,23) rectangle (56.2,27.2);
\node[rval, text=coffee, anchor=base] at (49.6,26) {14};
\fill[coffee] (56.2,23) rectangle (68.5,27.2);
\node[rval, text=lemon, anchor=base] at (62.4,26) {13};
\fill[taupe!55] (68.5,23) rectangle (72.4,27.2);
\fill[cinnamon] (72.4,23) rectangle (121.3,27.2);
\node[rval, text=lemon, anchor=base] at (96.9,26) {51};
\fill[cinnamon!62!lemon] (121.3,23) rectangle (138,27.2);
\node[rval, text=coffee, anchor=base] at (129.7,26) {17};
\fill[cinnamon!28!lemon] (138,23) rectangle (139,27.2);
\draw[taupe!45, line width=0.25pt] (43,23) rectangle (139,27.2);
\node[rtick, anchor=base east] at (41.5,31.4) {with listing};
\fill[coral] (43,28.3) rectangle (64.5,32.5);
\node[rval, text=coffee, anchor=base] at (53.8,31.3) {22};
\fill[coffee] (64.5,28.3) rectangle (72,32.5);
\node[rval, text=lemon, anchor=base] at (68.2,31.3) {8};
\fill[cinnamon] (72,28.3) rectangle (111.4,32.5);
\node[rval, text=lemon, anchor=base] at (91.7,31.3) {41};
\fill[cinnamon!62!lemon] (111.4,28.3) rectangle (113.8,32.5);
\fill[cinnamon!28!lemon] (113.8,28.3) rectangle (136.1,32.5);
\node[rval, text=coffee, anchor=base] at (125,31.3) {23};
\fill[lemon] (136.1,28.3) rectangle (139,32.5);
\draw[taupe!45, line width=0.25pt] (43,28.3) rectangle (139,32.5);
\node[rttl, anchor=base west] at (0,39.2) {Nemotron-3 Super};
\node[rtick, anchor=base east] at (41.5,39.1) {bits only};
\fill[coral] (43,36) rectangle (47.8,40.2);
\fill[coffee] (47.8,36) rectangle (134.2,40.2);
\node[rval, text=lemon, anchor=base] at (91,39) {90};
\fill[taupe!55] (134.2,36) rectangle (135.9,40.2);
\fill[cinnamon] (135.9,36) rectangle (138.2,40.2);
\fill[cinnamon!62!lemon] (138.2,36) rectangle (139,40.2);
\draw[taupe!45, line width=0.25pt] (43,36) rectangle (139,40.2);
\node[rtick, anchor=base east] at (41.5,44.4) {with listing};
\fill[coral] (43,41.3) rectangle (60.3,45.5);
\node[rval, text=coffee, anchor=base] at (51.6,44.3) {18};
\fill[coffee] (60.3,41.3) rectangle (125.9,45.5);
\node[rval, text=lemon, anchor=base] at (93.1,44.3) {68};
\fill[taupe!55] (125.9,41.3) rectangle (126.7,45.5);
\fill[cinnamon] (126.7,41.3) rectangle (134.4,45.5);
\node[rval, text=lemon, anchor=base] at (130.6,44.3) {8};
\fill[cinnamon!62!lemon] (134.4,41.3) rectangle (134.6,45.5);
\fill[cinnamon!28!lemon] (134.6,41.3) rectangle (138.6,45.5);
\fill[lemon] (138.6,41.3) rectangle (139,45.5);
\draw[taupe!45, line width=0.25pt] (43,41.3) rectangle (139,45.5);
\node[rttl, anchor=base west] at (0,52.2) {MiniMax-M3};
\node[rtick, anchor=base east] at (41.5,52.1) {bits only};
\fill[coral] (43,49) rectangle (52,53.2);
\node[rval, text=coffee, anchor=base] at (47.5,52) {9};
\fill[coffee] (52,49) rectangle (134.2,53.2);
\node[rval, text=lemon, anchor=base] at (93.1,52) {86};
\fill[taupe!55] (134.2,49) rectangle (134.6,53.2);
\fill[cinnamon] (134.6,49) rectangle (137.5,53.2);
\fill[cinnamon!62!lemon] (137.5,49) rectangle (138.2,53.2);
\fill[cinnamon!28!lemon] (138.2,49) rectangle (139,53.2);
\draw[taupe!45, line width=0.25pt] (43,49) rectangle (139,53.2);
\node[rtick, anchor=base east] at (41.5,57.4) {with listing};
\fill[coral] (43,54.3) rectangle (70.3,58.5);
\node[rval, text=coffee, anchor=base] at (56.6,57.3) {28};
\fill[coffee] (70.3,54.3) rectangle (115.2,58.5);
\node[rval, text=lemon, anchor=base] at (92.7,57.3) {47};
\fill[taupe!55] (115.2,54.3) rectangle (115.4,58.5);
\fill[cinnamon] (115.4,54.3) rectangle (131.5,58.5);
\node[rval, text=lemon, anchor=base] at (123.4,57.3) {17};
\fill[cinnamon!62!lemon] (131.5,54.3) rectangle (132.1,58.5);
\fill[cinnamon!28!lemon] (132.1,54.3) rectangle (137.7,58.5);
\fill[lemon] (137.7,54.3) rectangle (139,58.5);
\draw[taupe!45, line width=0.25pt] (43,54.3) rectangle (139,58.5);
\node[rttl, anchor=base west] at (0,65.2) {DeepSeek-V4.1-Flash};
\node[rtick, anchor=base east] at (41.5,65.1) {bits only};
\fill[coral] (43,62) rectangle (88.7,66.2);
\node[rval, text=coffee, anchor=base] at (65.8,65) {48};
\fill[coffee] (88.7,62) rectangle (137.5,66.2);
\node[rval, text=lemon, anchor=base] at (113.1,65) {51};
\fill[cinnamon] (137.5,62) rectangle (138.4,66.2);
\fill[cinnamon!28!lemon] (138.4,62) rectangle (139,66.2);
\draw[taupe!45, line width=0.25pt] (43,62) rectangle (139,66.2);
\node[rtick, anchor=base east] at (41.5,70.4) {with listing};
\fill[coral] (43,67.3) rectangle (137.1,71.5);
\node[rval, text=coffee, anchor=base] at (90,70.3) {98};
\fill[cinnamon] (137.1,67.3) rectangle (137.8,71.5);
\fill[cinnamon!62!lemon] (137.8,67.3) rectangle (138,71.5);
\fill[cinnamon!28!lemon] (138,67.3) rectangle (139,71.5);
\draw[taupe!45, line width=0.25pt] (43,67.3) rectangle (139,71.5);
\node[rttl, anchor=base west] at (0,78.2) {Claude Opus 5.5};
\node[rtick, anchor=base east] at (41.5,78.1) {bits only};
\fill[coral] (43,75) rectangle (137.3,79.2);
\node[rval, text=coffee, anchor=base] at (90.1,78) {98};
\fill[cinnamon] (137.3,75) rectangle (137.5,79.2);
\fill[cinnamon!28!lemon] (137.5,75) rectangle (139,79.2);
\draw[taupe!45, line width=0.25pt] (43,75) rectangle (139,79.2);
\node[rtick, anchor=base east] at (41.5,83.4) {with listing};
\fill[coral] (43,80.3) rectangle (138.6,84.5);
\node[rval, text=coffee, anchor=base] at (90.8,83.3) {100};
\fill[cinnamon!28!lemon] (138.6,80.3) rectangle (139,84.5);
\draw[taupe!45, line width=0.25pt] (43,80.3) rectangle (139,84.5);
\node[rtick, anchor=base east] at (139,89) {numbers inside the bars are percentages of the runs of that row};
\end{tikzpicture}
  }
  \caption{Outcome of every run, by the four checks of the scoring procedure. Each bar is the 500 runs of one model in one condition.}
  \Description{Two stacked bars per model, one for the bits only and one with the disassembler listing, split into solved, cut, no edit, not applicable, wrong place, wrong change and broke the rest.}
  \label{fig:outcomes-results}
\end{figure}

\paragraph{The decoding bottleneck}
The listing turns the bits into instructions, with their addresses and the sections of the file; the executable in the prompt is the same. The gain from the listing is therefore our measure of the decoding bottleneck defined in the Introduction, and Section~\ref{sec:threats} states what else this gain includes. For DeepSeek-V4.1-Flash, little remains after decoding: with the listing it solves 490 of 500 runs. Claude Opus 5.5 shows little decoding bottleneck: with the bits only it solves 491 of 500 runs. For the four models, the second bottleneck remains: with the listing, where the instructions are given, they leave most of their runs unsolved. Figure~\ref{fig:outcomes-results} makes this visible: with the listing, most of their unsolved runs reach the output limit or give an edit that cannot be applied, a pattern that Section~\ref{sec:rq3} examines step by step. The gain does not come from the output limit alone. When all models are held to the 32,768 output tokens at which the provider stops Nemotron-3 Super, a comparison that Supplementary Table~S3 gives in full, every model still solves more runs with the listing than with the bits.

\paragraph{Decoding costs output tokens}
The last two columns of Table~\ref{tab:models} give the mean output tokens per run: with the listing, DeepSeek-V4.1-Flash produces about 83 percent fewer. With the bits only, about half of its runs reach the output limit, and none do with the listing. Figure~\ref{fig:bottleneck} shows one task of the study, which the same model solves with far fewer tokens when it has the listing. Qwen3-235B, Nemotron-3 Super and MiniMax-M3 also produce fewer tokens with the listing. gpt-oss-120b is the exception: it is never cut, and its output grows. Claude Opus 5.5 produces the fewest output tokens of all models, 1,363 per run with the bits only and 1,058 with the listing, and none of its runs is cut.

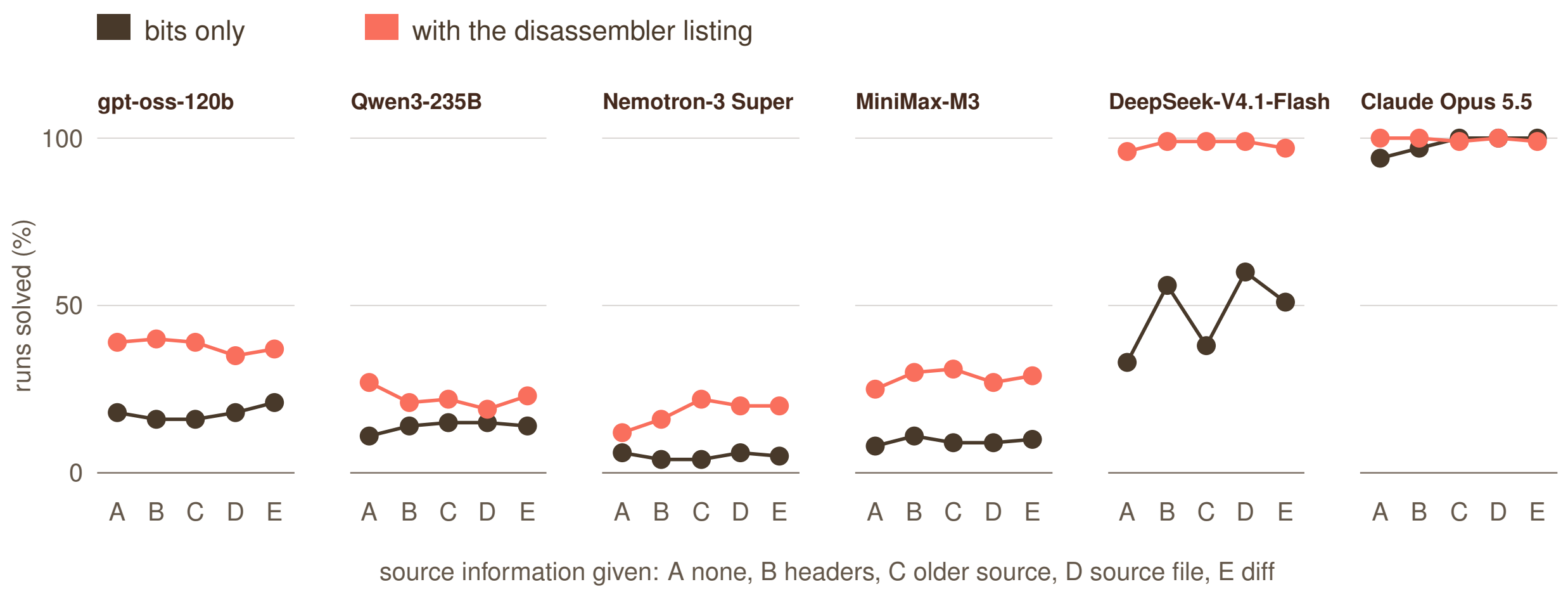
\begin{figure}[t]
  \centering
  \resizebox{\linewidth}{!}{
\begin{tikzpicture}[x=1mm, y=-1mm]
\useasboundingbox (0,0) rectangle (139,53.5);
\fill[taupe] (8,0.9000000000000004) rectangle (11,3.2);
\node[rlab, anchor=base west] at (12.2,3.2) {bits only};
\fill[coral] (32,0.9000000000000004) rectangle (35,3.2);
\node[rlab, anchor=base west] at (36.2,3.2) {with the disassembler listing};
\node[rttl, anchor=base west, scale=0.7] at (8,9.4) {gpt-oss-120b};
\draw[rgrid] (8,42) -- (25.7,42);
\node[rtick, anchor=east] at (6.8,42) {0};
\draw[rgrid] (8,27) -- (25.7,27);
\node[rtick, anchor=east] at (6.8,27) {50};
\draw[rgrid] (8,12) -- (25.7,12);
\node[rtick, anchor=east] at (6.8,12) {100};
\draw[raxis] (8,42) -- (25.7,42);
\node[rtick, anchor=base] at (9.8,46.3) {A};
\node[rtick, anchor=base] at (13.3,46.3) {B};
\node[rtick, anchor=base] at (16.8,46.3) {C};
\node[rtick, anchor=base] at (20.4,46.3) {D};
\node[rtick, anchor=base] at (23.9,46.3) {E};
\draw[taupe, line width=0.9pt, line join=round] (9.8,36.6) -- (13.3,37.2) -- (16.8,37.2) -- (20.4,36.6) -- (23.9,35.7);
\fill[taupe] (9.8,36.6) circle (0.85);
\fill[taupe] (13.3,37.2) circle (0.85);
\fill[taupe] (16.8,37.2) circle (0.85);
\fill[taupe] (20.4,36.6) circle (0.85);
\fill[taupe] (23.9,35.7) circle (0.85);
\draw[coral, line width=0.9pt, line join=round] (9.8,30.3) -- (13.3,30) -- (16.8,30.3) -- (20.4,31.5) -- (23.9,30.9);
\fill[coral] (9.8,30.3) circle (0.85);
\fill[coral] (13.3,30) circle (0.85);
\fill[coral] (16.8,30.3) circle (0.85);
\fill[coral] (20.4,31.5) circle (0.85);
\fill[coral] (23.9,30.9) circle (0.85);
\node[rttl, anchor=base west, scale=0.7] at (30.7,9.4) {Qwen3-235B};
\draw[rgrid] (30.7,42) -- (48.3,42);
\draw[rgrid] (30.7,27) -- (48.3,27);
\draw[rgrid] (30.7,12) -- (48.3,12);
\draw[raxis] (30.7,42) -- (48.3,42);
\node[rtick, anchor=base] at (32.4,46.3) {A};
\node[rtick, anchor=base] at (36,46.3) {B};
\node[rtick, anchor=base] at (39.5,46.3) {C};
\node[rtick, anchor=base] at (43,46.3) {D};
\node[rtick, anchor=base] at (46.6,46.3) {E};
\draw[taupe, line width=0.9pt, line join=round] (32.4,38.7) -- (36,37.8) -- (39.5,37.5) -- (43,37.5) -- (46.6,37.8);
\fill[taupe] (32.4,38.7) circle (0.85);
\fill[taupe] (36,37.8) circle (0.85);
\fill[taupe] (39.5,37.5) circle (0.85);
\fill[taupe] (43,37.5) circle (0.85);
\fill[taupe] (46.6,37.8) circle (0.85);
\draw[coral, line width=0.9pt, line join=round] (32.4,33.9) -- (36,35.7) -- (39.5,35.4) -- (43,36.3) -- (46.6,35.1);
\fill[coral] (32.4,33.9) circle (0.85);
\fill[coral] (36,35.7) circle (0.85);
\fill[coral] (39.5,35.4) circle (0.85);
\fill[coral] (43,36.3) circle (0.85);
\fill[coral] (46.6,35.1) circle (0.85);
\node[rttl, anchor=base west, scale=0.7] at (53.3,9.4) {Nemotron-3 Super};
\draw[rgrid] (53.3,42) -- (71,42);
\draw[rgrid] (53.3,27) -- (71,27);
\draw[rgrid] (53.3,12) -- (71,12);
\draw[raxis] (53.3,42) -- (71,42);
\node[rtick, anchor=base] at (55.1,46.3) {A};
\node[rtick, anchor=base] at (58.6,46.3) {B};
\node[rtick, anchor=base] at (62.2,46.3) {C};
\node[rtick, anchor=base] at (65.7,46.3) {D};
\node[rtick, anchor=base] at (69.2,46.3) {E};
\draw[taupe, line width=0.9pt, line join=round] (55.1,40.2) -- (58.6,40.8) -- (62.2,40.8) -- (65.7,40.2) -- (69.2,40.5);
\fill[taupe] (55.1,40.2) circle (0.85);
\fill[taupe] (58.6,40.8) circle (0.85);
\fill[taupe] (62.2,40.8) circle (0.85);
\fill[taupe] (65.7,40.2) circle (0.85);
\fill[taupe] (69.2,40.5) circle (0.85);
\draw[coral, line width=0.9pt, line join=round] (55.1,38.4) -- (58.6,37.2) -- (62.2,35.4) -- (65.7,36) -- (69.2,36);
\fill[coral] (55.1,38.4) circle (0.85);
\fill[coral] (58.6,37.2) circle (0.85);
\fill[coral] (62.2,35.4) circle (0.85);
\fill[coral] (65.7,36) circle (0.85);
\fill[coral] (69.2,36) circle (0.85);
\node[rttl, anchor=base west, scale=0.7] at (76,9.4) {MiniMax-M3};
\draw[rgrid] (76,42) -- (93.7,42);
\draw[rgrid] (76,27) -- (93.7,27);
\draw[rgrid] (76,12) -- (93.7,12);
\draw[raxis] (76,42) -- (93.7,42);
\node[rtick, anchor=base] at (77.8,46.3) {A};
\node[rtick, anchor=base] at (81.3,46.3) {B};
\node[rtick, anchor=base] at (84.8,46.3) {C};
\node[rtick, anchor=base] at (88.4,46.3) {D};
\node[rtick, anchor=base] at (91.9,46.3) {E};
\draw[taupe, line width=0.9pt, line join=round] (77.8,39.6) -- (81.3,38.7) -- (84.8,39.3) -- (88.4,39.3) -- (91.9,39);
\fill[taupe] (77.8,39.6) circle (0.85);
\fill[taupe] (81.3,38.7) circle (0.85);
\fill[taupe] (84.8,39.3) circle (0.85);
\fill[taupe] (88.4,39.3) circle (0.85);
\fill[taupe] (91.9,39) circle (0.85);
\draw[coral, line width=0.9pt, line join=round] (77.8,34.5) -- (81.3,33) -- (84.8,32.7) -- (88.4,33.9) -- (91.9,33.3);
\fill[coral] (77.8,34.5) circle (0.85);
\fill[coral] (81.3,33) circle (0.85);
\fill[coral] (84.8,32.7) circle (0.85);
\fill[coral] (88.4,33.9) circle (0.85);
\fill[coral] (91.9,33.3) circle (0.85);
\node[rttl, anchor=base west, scale=0.7] at (98.7,9.4) {DeepSeek-V4.1-Flash};
\draw[rgrid] (98.7,42) -- (116.3,42);
\draw[rgrid] (98.7,27) -- (116.3,27);
\draw[rgrid] (98.7,12) -- (116.3,12);
\draw[raxis] (98.7,42) -- (116.3,42);
\node[rtick, anchor=base] at (100.4,46.3) {A};
\node[rtick, anchor=base] at (104,46.3) {B};
\node[rtick, anchor=base] at (107.5,46.3) {C};
\node[rtick, anchor=base] at (111,46.3) {D};
\node[rtick, anchor=base] at (114.6,46.3) {E};
\draw[taupe, line width=0.9pt, line join=round] (100.4,32.1) -- (104,25.2) -- (107.5,30.6) -- (111,24) -- (114.6,26.7);
\fill[taupe] (100.4,32.1) circle (0.85);
\fill[taupe] (104,25.2) circle (0.85);
\fill[taupe] (107.5,30.6) circle (0.85);
\fill[taupe] (111,24) circle (0.85);
\fill[taupe] (114.6,26.7) circle (0.85);
\draw[coral, line width=0.9pt, line join=round] (100.4,13.2) -- (104,12.3) -- (107.5,12.3) -- (111,12.3) -- (114.6,12.9);
\fill[coral] (100.4,13.2) circle (0.85);
\fill[coral] (104,12.3) circle (0.85);
\fill[coral] (107.5,12.3) circle (0.85);
\fill[coral] (111,12.3) circle (0.85);
\fill[coral] (114.6,12.9) circle (0.85);
\node[rttl, anchor=base west, scale=0.7] at (121.3,9.4) {Claude Opus 5.5};
\draw[rgrid] (121.3,42) -- (139,42);
\draw[rgrid] (121.3,27) -- (139,27);
\draw[rgrid] (121.3,12) -- (139,12);
\draw[raxis] (121.3,42) -- (139,42);
\node[rtick, anchor=base] at (123.1,46.3) {A};
\node[rtick, anchor=base] at (126.6,46.3) {B};
\node[rtick, anchor=base] at (130.2,46.3) {C};
\node[rtick, anchor=base] at (133.7,46.3) {D};
\node[rtick, anchor=base] at (137.2,46.3) {E};
\draw[taupe, line width=0.9pt, line join=round] (123.1,13.8) -- (126.6,12.9) -- (130.2,12) -- (133.7,12) -- (137.2,12);
\fill[taupe] (123.1,13.8) circle (0.85);
\fill[taupe] (126.6,12.9) circle (0.85);
\fill[taupe] (130.2,12) circle (0.85);
\fill[taupe] (133.7,12) circle (0.85);
\fill[taupe] (137.2,12) circle (0.85);
\draw[coral, line width=0.9pt, line join=round] (123.1,12) -- (126.6,12) -- (130.2,12.3) -- (133.7,12) -- (137.2,12.3);
\fill[coral] (123.1,12) circle (0.85);
\fill[coral] (126.6,12) circle (0.85);
\fill[coral] (130.2,12.3) circle (0.85);
\fill[coral] (133.7,12) circle (0.85);
\fill[coral] (137.2,12.3) circle (0.85);
\node[rtick, anchor=base, rotate=90] at (2.0,27.0) {runs solved (\%)};
\node[rtick, anchor=base] at (73.5,51.6) {source information given: A none, B headers, C older source, D source file, E diff};
\end{tikzpicture}
  }
  \caption{Runs solved by source information, for each model, with the bits only and with the disassembler listing. Each point is 100 runs. Levels: A none, B headers, C older source, D source file, E diff.}
  \Description{One panel per model. Each panel has two lines over the five levels of source information: runs solved with the bits only and with the disassembler listing.}
  \label{fig:levels}
\end{figure}

\paragraph{Source information changes the result for two models, with the bits only}
The second factor of RQ1 is the level of source information. Figure~\ref{fig:levels} follows each model across the five levels. The pattern of Table~\ref{tab:models} holds at level A, with no source information: Claude Opus 5.5 solves 94 of its 100 runs with the bits only, DeepSeek-V4.1-Flash goes from 33 to 96 with the listing, and each of the other four solves at most 39 with it. For DeepSeek-V4.1-Flash with the bits only, the levels differ, and the test of Table~\ref{tab:tests} gives an adjusted $p < 0.001$. Level B, which gives declarations with their one-line comments and no code, comes close to the complete source file at level D. The exact diff at level E gives fewer runs solved than level D. The widest spread of any model, 27 points between levels A and D, is about half the 50-point gain that the listing gives the same model. Supplementary Table~S4 counts the runs of this condition that finish within the output limit: almost every one of them is solved, at every level. What differs between levels is how many runs finish within the limit. Claude Opus 5.5 with the bits only also differs between levels (adjusted $p = 0.009$): it solves 94 runs at level A, 97 at level B and all 100 at levels C, D and E.

\paragraph{For the other conditions, the tests detect no difference between levels}
For the four other models, and for DeepSeek-V4.1-Flash and Claude Opus 5.5 with the listing, no adjusted $p$-value in Table~\ref{tab:tests} reaches significance. With the listing, DeepSeek-V4.1-Flash and Claude Opus 5.5 solve almost every run at every level, so there is little room to differ. In Figure~\ref{fig:levels}, the other four models differ by a few points between levels, the widest being Nemotron-3 Super with the listing. With 100 tasks, the test does not separate such differences from chance, and the intervals that Supplementary Table~S5 gives for the paired differences between levels are wide. For these conditions the data show no effect that the design can detect; they do not show that source information has no effect.

\subsection{RQ2: Kind of change}
\label{sec:rq2}

RQ1 shows how much the listing and the source information change the runs solved; RQ2 asks which changes are solved. The answer depends on whether the edit touches data or instructions, and on the kind of change.

\paragraph{With the bits only, four models solve edits to data and rarely edits to instructions}
Table~\ref{tab:families} separates the 15 data tasks from the 85 instruction tasks, and Figure~\ref{fig:kinds} counts the runs solved in each group. It has two grids, one for the bits only and one with the listing, and each row is a model. The first column holds the data tasks (families V3, V5 and D1), and the other three split the instruction tasks by kind of change. With the bits only, the four models solve far more of the data runs than of the instruction runs, and few of the instruction runs at all. DeepSeek-V4.1-Flash shows the same difference at a higher level: it solves almost all data runs and about two fifths of the instruction runs. Claude Opus 5.5 solves all 75 data runs and 416 of the 425 instruction runs. The tasks of a family are variants of one program, so the 20 families are the independent units of this comparison. Supplementary Table~S6 repeats it with the family as the unit: every model solves the data families more often than the instruction families, Claude Opus 5.5 by 2 points.

\begin{figure}[t]
  \centering
  \resizebox{\linewidth}{!}{
\begin{tikzpicture}[x=1mm, y=-1mm]
\useasboundingbox (0,0) rectangle (139,60.400000000000006);
\node[rttl, anchor=base west] at (34,3.4) {bits only};
\node[rtick, anchor=base west] at (46.7,7.4) {edits to instructions};
\draw[taupe!45, line width=0.3pt] (46.7,8.4) -- (79.4,8.4);
\node[rtick, anchor=base] at (39.8,11.6) {data only};
\node[rtick, anchor=base] at (51.2,11.6) {value};
\node[rtick, anchor=base] at (62.8,11.6) {decision};
\node[rtick, anchor=base] at (74.2,11.6) {insert logic};
\node[rttl, anchor=base west] at (0,17.2) {gpt-oss-120b};
\foreach \kv/\kn/\kp [count=\kk from 0] in {63/75/84, 6/100/6, 20/150/13, 0/175/0} { \fill[coral!\kp!lemon] ({34.5+11.5*\kk},13) rectangle ({45+11.5*\kk},19.2); \node[rval, text=coffee, anchor=base] at ({39.8+11.5*\kk},17.1) {\kv/\kn}; }
\node[rttl, anchor=base west] at (0,24.6) {Qwen3-235B};
\foreach \kv/\kn/\kp [count=\kk from 0] in {63/75/84, 1/100/1, 5/150/3, 0/175/0} { \fill[coral!\kp!lemon] ({34.5+11.5*\kk},20.4) rectangle ({45+11.5*\kk},26.6); \node[rval, text=coffee, anchor=base] at ({39.8+11.5*\kk},24.5) {\kv/\kn}; }
\node[rttl, anchor=base west] at (0,32) {Nemotron-3 Super};
\foreach \kv/\kn/\kp [count=\kk from 0] in {22/75/29, 0/100/0, 3/150/2, 0/175/0} { \fill[coral!\kp!lemon] ({34.5+11.5*\kk},27.8) rectangle ({45+11.5*\kk},34); \node[rval, text=coffee, anchor=base] at ({39.8+11.5*\kk},31.9) {\kv/\kn}; }
\node[rttl, anchor=base west] at (0,39.4) {MiniMax-M3};
\foreach \kv/\kn/\kp [count=\kk from 0] in {41/75/55, 4/100/4, 1/150/1, 1/175/1} { \fill[coral!\kp!lemon] ({34.5+11.5*\kk},35.2) rectangle ({45+11.5*\kk},41.4); \node[rval, text=coffee, anchor=base] at ({39.8+11.5*\kk},39.3) {\kv/\kn}; }
\node[rttl, anchor=base west] at (0,46.8) {DeepSeek-V4.1-Flash};
\foreach \kv/\kn/\kp [count=\kk from 0] in {72/75/96, 30/100/30, 69/150/46, 67/175/38} { \fill[coral!\kp!lemon] ({34.5+11.5*\kk},42.6) rectangle ({45+11.5*\kk},48.8); \node[rval, text=coffee, anchor=base] at ({39.8+11.5*\kk},46.7) {\kv/\kn}; }
\node[rttl, anchor=base west] at (0,54.2) {Claude Opus 5.5};
\foreach \kv/\kn/\kp [count=\kk from 0] in {75/75/100, 99/100/99, 150/150/100, 167/175/95} { \fill[coral!\kp!lemon] ({34.5+11.5*\kk},50) rectangle ({45+11.5*\kk},56.2); \node[rval, text=coffee, anchor=base] at ({39.8+11.5*\kk},54.1) {\kv/\kn}; }
\node[rttl, anchor=base west] at (89,3.4) {with the disassembler listing};
\node[rtick, anchor=base west] at (101.7,7.4) {edits to instructions};
\draw[taupe!45, line width=0.3pt] (101.7,8.4) -- (134.4,8.4);
\node[rtick, anchor=base] at (94.8,11.6) {data only};
\node[rtick, anchor=base] at (106.2,11.6) {value};
\node[rtick, anchor=base] at (117.8,11.6) {decision};
\node[rtick, anchor=base] at (129.2,11.6) {insert logic};
\foreach \kv/\kn/\kp [count=\kk from 0] in {70/75/93, 53/100/53, 61/150/41, 6/175/3} { \fill[coral!\kp!lemon] ({89.5+11.5*\kk},13) rectangle ({100+11.5*\kk},19.2); \node[rval, text=coffee, anchor=base] at ({94.8+11.5*\kk},17.1) {\kv/\kn}; }
\foreach \kv/\kn/\kp [count=\kk from 0] in {58/75/77, 19/100/19, 34/150/23, 1/175/1} { \fill[coral!\kp!lemon] ({89.5+11.5*\kk},20.4) rectangle ({100+11.5*\kk},26.6); \node[rval, text=coffee, anchor=base] at ({94.8+11.5*\kk},24.5) {\kv/\kn}; }
\foreach \kv/\kn/\kp [count=\kk from 0] in {51/75/68, 17/100/17, 21/150/14, 1/175/1} { \fill[coral!\kp!lemon] ({89.5+11.5*\kk},27.8) rectangle ({100+11.5*\kk},34); \node[rval, text=coffee, anchor=base] at ({94.8+11.5*\kk},31.9) {\kv/\kn}; }
\foreach \kv/\kn/\kp [count=\kk from 0] in {47/75/63, 34/100/34, 50/150/33, 11/175/6} { \fill[coral!\kp!lemon] ({89.5+11.5*\kk},35.2) rectangle ({100+11.5*\kk},41.4); \node[rval, text=coffee, anchor=base] at ({94.8+11.5*\kk},39.3) {\kv/\kn}; }
\foreach \kv/\kn/\kp [count=\kk from 0] in {74/75/99, 96/100/96, 149/150/99, 171/175/98} { \fill[coral!\kp!lemon] ({89.5+11.5*\kk},42.6) rectangle ({100+11.5*\kk},48.8); \node[rval, text=coffee, anchor=base] at ({94.8+11.5*\kk},46.7) {\kv/\kn}; }
\foreach \kv/\kn/\kp [count=\kk from 0] in {75/75/100, 100/100/100, 148/150/99, 175/175/100} { \fill[coral!\kp!lemon] ({89.5+11.5*\kk},50) rectangle ({100+11.5*\kk},56.2); \node[rval, text=coffee, anchor=base] at ({94.8+11.5*\kk},54.1) {\kv/\kn}; }
\end{tikzpicture}
  }
  \caption{Runs solved by what the edit touches. \emph{Data only}: the reference solution changes a text or a table of values. Edits to instructions are split by kind of change. Each cell gives solved runs over runs, over the five levels of source information; darker cells are higher shares.}
  \Description{Two grids of cells, one for the bits only and one with the disassembler listing. Rows are models. Columns are data-only edits and three kinds of edits to instructions: value, decision and insert logic.}
  \label{fig:kinds}
\end{figure}

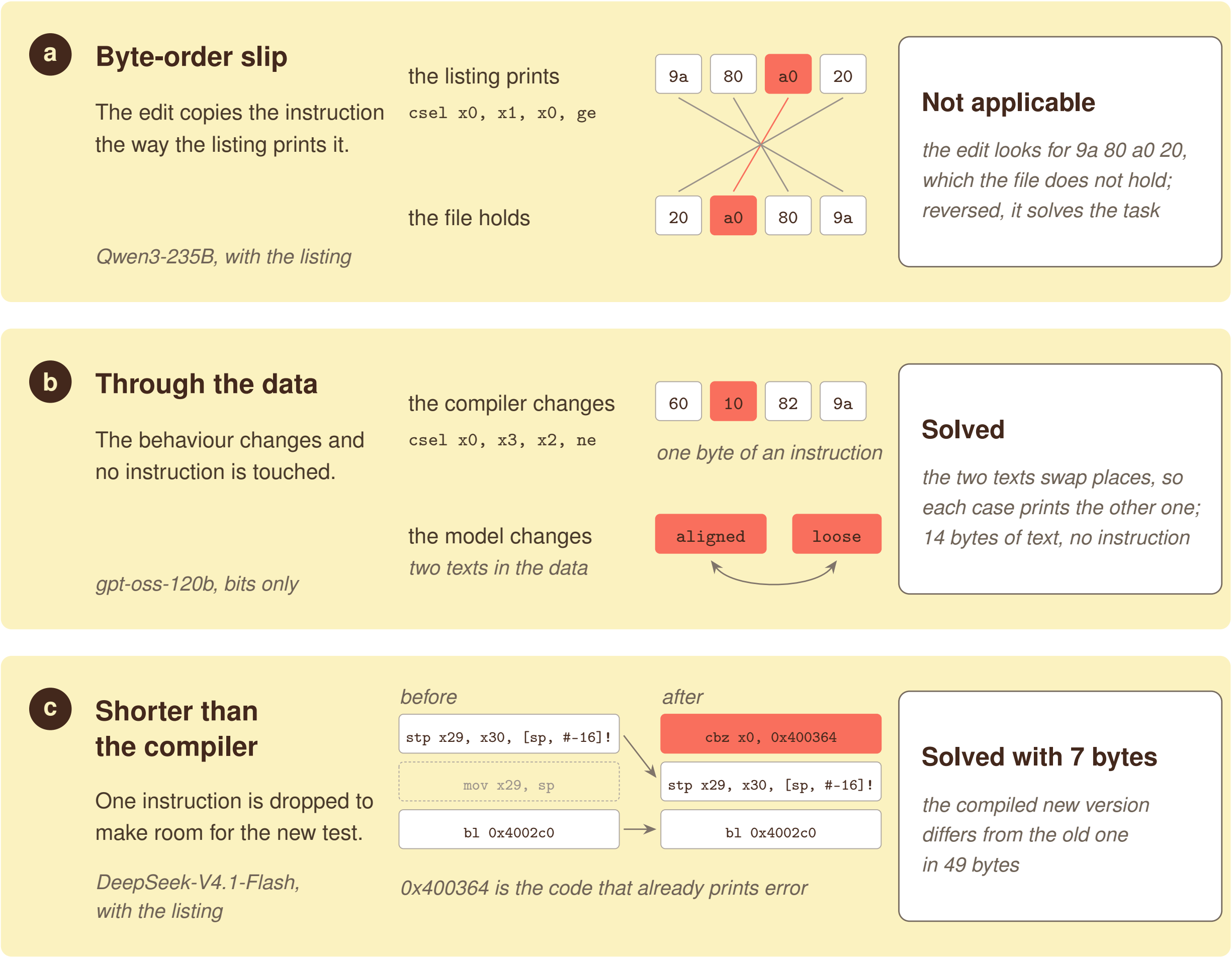
\begin{figure}[t]
  \centering
  \resizebox{\linewidth}{!}{
\begin{tikzpicture}[x=1mm, y=-1mm,
  bt/.style={anchor=base west, inner sep=0, text=coffee, font=\fontfamily{phv}\fontsize{8.8}{10.5}\selectfont\bfseries},
  bd/.style={anchor=base west, inner sep=0, text=taupe, font=\fontfamily{phv}\fontsize{7}{8.8}\selectfont},
  bi/.style={anchor=base west, inner sep=0, text=taupe!80, font=\fontfamily{phv}\fontsize{6.5}{8}\selectfont\itshape},
  bm/.style={anchor=base west, inner sep=0, text=taupe, font=\ttfamily\fontsize{6.2}{7.5}\selectfont},
  bx/.style={anchor=base, inner sep=0, text=coffee, font=\ttfamily\fontsize{6.5}{7.5}\selectfont},
  bs/.style={anchor=base, inner sep=0, text=coffee, font=\ttfamily\fontsize{5}{6.5}\selectfont},
  bnm/.style={anchor=base west, inner sep=0, text=coffee, font=\fontfamily{phv}\fontsize{8.2}{10}\selectfont\bfseries},
  bnum/.style={circle, fill=coffee, inner sep=0, minimum size=4.8mm, text=lemon, font=\fontfamily{phv}\fontsize{8}{8.8}\selectfont\bfseries},
  bchip/.style={draw=taupe!75, line width=0.45pt, rounded corners=1.2mm, fill=white},
  bcell/.style={draw=taupe!45, line width=0.3pt, rounded corners=0.5mm, fill=white},
  bhot/.style={draw=coral, line width=0.3pt, rounded corners=0.5mm, fill=coral},
  bmove/.style={draw=taupe!70, line width=0.45pt, -{Stealth[length=1.5mm, width=1.2mm]}}]
\useasboundingbox (0,0) rectangle (139,108);
\foreach \y in {0,37,74} \fill[lemon, rounded corners=1.2mm] (0,\y) rectangle (139,\y+34);
\node[bnum] at (5.6,6) {a};
\node[bt] at (10.6,7.3) {Byte-order slip};
\node[bd] at (10.6,13.4) {The edit copies the instruction};
\node[bd] at (10.6,17) {the way the listing prints it.};
\node[bi] at (10.6,29.6) {Qwen3-235B, with the listing};
\node[bd] at (46,9.3) {the listing prints};
\node[bm] at (46,13.2) {csel x0, x1, x0, ge};
\node[bd] at (46,25.3) {the file holds};
\foreach \k/\a/\b in {0/3/taupe!55, 1/2/taupe!55, 2/1/coral, 3/0/taupe!55}
  \draw[\b, line width=0.45pt] (76.6+6.2*\k,10.9) -- (76.6+6.2*\a,21.5);
\foreach \k/\t/\s in {0/9a/bcell, 1/80/bcell, 2/a0/bhot, 3/20/bcell} {
  \draw[\s] (74+6.2*\k,6) rectangle (79.2+6.2*\k,10.4); \node[bx] at (76.6+6.2*\k,9.1) {\t}; }
\foreach \k/\t/\s in {0/20/bcell, 1/a0/bhot, 2/80/bcell, 3/9a/bcell} {
  \draw[\s] (74+6.2*\k,22) rectangle (79.2+6.2*\k,26.4); \node[bx] at (76.6+6.2*\k,25.1) {\t}; }
\draw[bchip] (101.5,4) rectangle (138,30);
\node[bnm] at (104,12.4) {Not applicable};
\node[bi] at (104,17.6) {the edit looks for 9a 80 a0 20,};
\node[bi] at (104,21) {which the file does not hold;};
\node[bi] at (104,24.4) {reversed, it solves the task};
\node[bnum] at (5.6,43) {b};
\node[bt] at (10.6,44.3) {Through the data};
\node[bd] at (10.6,50.4) {The behaviour changes and};
\node[bd] at (10.6,54) {no instruction is touched.};
\node[bi] at (10.6,66.6) {gpt-oss-120b, bits only};
\node[bd] at (46,46.3) {the compiler changes};
\node[bm] at (46,50.2) {csel x0, x3, x2, ne};
\node[bd] at (46,61.3) {the model changes};
\node[bi] at (46,64.8) {two texts in the data};
\foreach \k/\t/\s in {0/60/bcell, 1/10/bhot, 2/82/bcell, 3/9a/bcell} {
  \draw[\s] (74+6.2*\k,43) rectangle (79.2+6.2*\k,47.4); \node[bx] at (76.6+6.2*\k,46.1) {\t}; }
\node[bi, anchor=base west] at (74,51.8) {one byte of an instruction};
\draw[bhot] (74,58) rectangle (86.5,62.4); \node[bx] at (80.25,61.1) {aligned};
\draw[bhot] (89.5,58) rectangle (99.5,62.4); \node[bx] at (94.5,61.1) {loose};
\draw[taupe!70, line width=0.45pt, {Stealth[length=1.5mm, width=1.2mm]}-{Stealth[length=1.5mm, width=1.2mm]}] (80.25,63.2) .. controls (82.75,66.6) and (92,66.6) .. (94.5,63.2);
\draw[bchip] (101.5,41) rectangle (138,67);
\node[bnm] at (104,49.4) {Solved};
\node[bi] at (104,54.6) {the two texts swap places, so};
\node[bi] at (104,58) {each case prints the other one;};
\node[bi] at (104,61.4) {14 bytes of text, no instruction};
\node[bnum] at (5.6,80) {c};
\node[bt] at (10.6,81.3) {Shorter than};
\node[bt] at (10.6,85.3) {the compiler};
\node[bd] at (10.6,91.2) {One instruction is dropped to};
\node[bd] at (10.6,94.8) {make room for the new test.};
\node[bi] at (10.6,100.4) {DeepSeek-V4.1-Flash,};
\node[bi] at (10.6,103.6) {with the listing};
\node[bi] at (45,79.4) {before};
\node[bi] at (74.6,79.4) {after};
\draw[bcell] (45,80.6) rectangle (69.9,85); \node[bs] at (57.45,83.7) {stp x29, x30, [sp, \#-16]!};
\draw[bhot] (74.6,80.6) rectangle (99.5,85); \node[bs] at (87.05,83.7) {cbz x0, 0x400364};
\draw[taupe!45, line width=0.3pt, rounded corners=0.5mm, dash pattern=on 0.8pt off 0.6pt] (45,86) rectangle (69.9,90.4); \node[bs, text=taupe!55] at (57.45,89.1) {mov x29, sp};
\draw[bcell] (74.6,86) rectangle (99.5,90.4); \node[bs] at (87.05,89.1) {stp x29, x30, [sp, \#-16]!};
\draw[bcell] (45,91.4) rectangle (69.9,95.8); \node[bs] at (57.45,94.5) {bl 0x4002c0};
\draw[bcell] (74.6,91.4) rectangle (99.5,95.8); \node[bs] at (87.05,94.5) {bl 0x4002c0};
\draw[bmove] (70.4,83) -- (74.1,87.8);
\draw[bmove] (70.4,93.6) -- (74.1,93.6);
\node[bi] at (45,101) {0x400364 is the code that already prints error};
\draw[bchip] (101.5,78) rectangle (138,104);
\node[bnm] at (104,86.4) {Solved with 7 bytes};
\node[bi] at (104,91.6) {the compiled new version};
\node[bi] at (104,95) {differs from the old one};
\node[bi] at (104,98.4) {in 49 bytes};
\end{tikzpicture}
  }
  \caption{Three behaviours seen in the runs, each drawn from one stored run with its real bytes. (a) Byte-order slip: the disassembler prints each instruction as one number, and the edit searches the file for the bytes in that order. (b) Through the data: asked to negate a test, the edit swaps the two texts that the program prints and leaves the instruction as it is. (c) Shorter than the compiler: the edit removes the instruction that sets the frame pointer and puts in its place a test that jumps to code already in the program.}
  \Description{Three rows. Row a shows the four bytes of an instruction as the listing prints them and, below, the same bytes in the reverse order in which the file holds them. Row b shows the one byte that the compiler changes and the two texts that the model swaps. Row c shows three instructions before and after the edit: one is removed, one moves down and a new test takes the first place.}
  \label{fig:behaviours}
\end{figure}

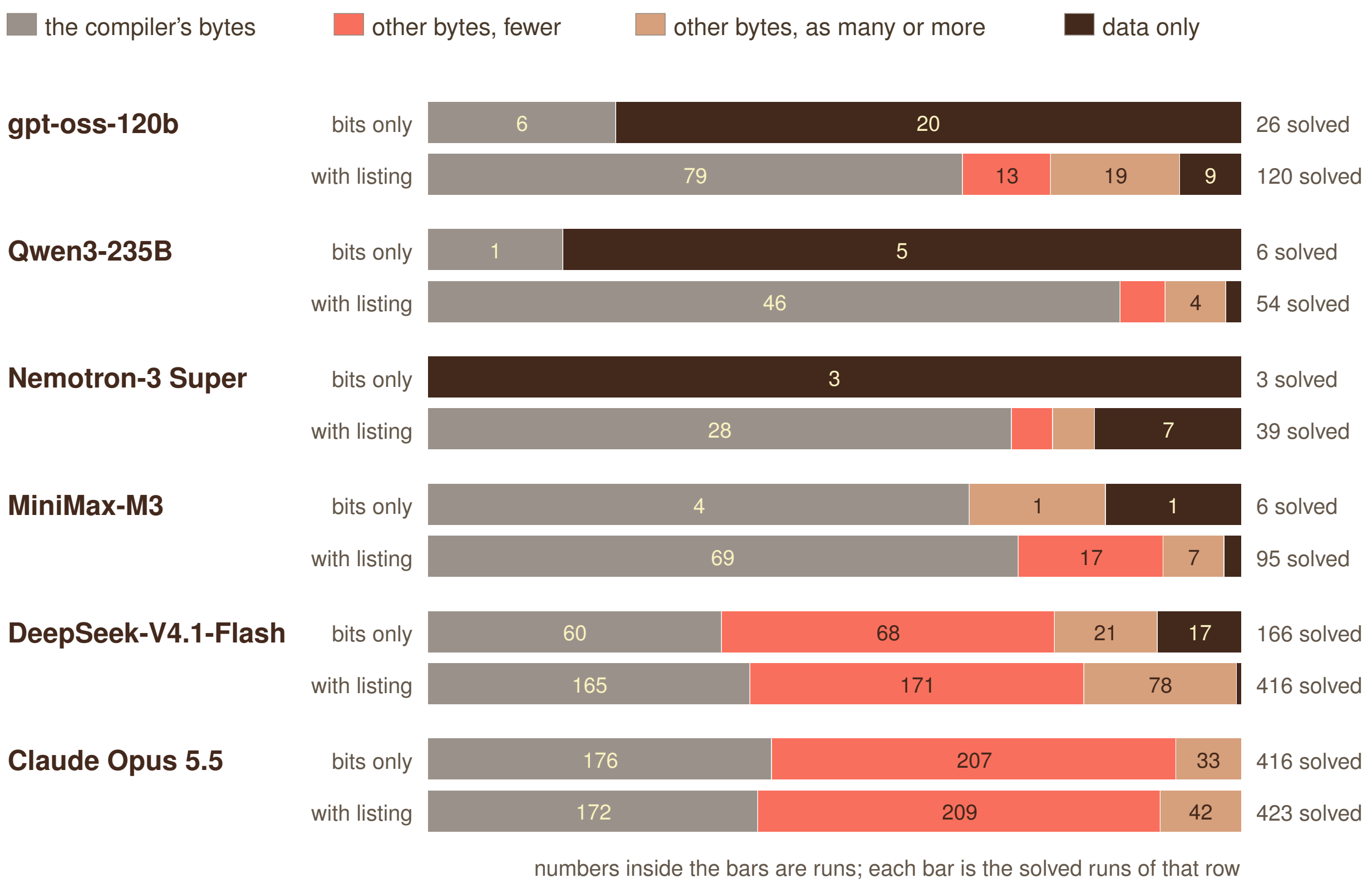
\begin{figure}[t]
  \centering
  \resizebox{\linewidth}{!}{
\begin{tikzpicture}[x=1mm, y=-1mm]
\useasboundingbox (0,0) rectangle (139,94);
\draw[taupe!60, line width=0.25pt, fill=taupe!55] (0,1.0) rectangle (3,3.2);
\node[rlab, anchor=base west] at (3.8,3.2) {the compiler's bytes};
\draw[taupe!60, line width=0.25pt, fill=coral] (33.4,1.0) rectangle (36.4,3.2);
\node[rlab, anchor=base west] at (37.2,3.2) {other bytes, fewer};
\draw[taupe!60, line width=0.25pt, fill=cinnamon!62!lemon] (64.2,1.0) rectangle (67.2,3.2);
\node[rlab, anchor=base west] at (68,3.2) {other bytes, as many or more};
\draw[taupe!60, line width=0.25pt, fill=coffee] (108,1.0) rectangle (111,3.2);
\node[rlab, anchor=base west] at (111.8,3.2) {data only};
\node[rttl, anchor=base west] at (0,13.2) {gpt-oss-120b};
\node[rtick, anchor=base east] at (41.5,13.1) {bits only};
\fill[taupe!55] (43,10) rectangle (62.2,14.2);
\node[rval, text=lemon, anchor=base] at (52.6,13) {6};
\draw[white, line width=0.5pt] (62.2,10) -- (62.2,14.2);
\fill[coffee] (62.2,10) rectangle (126,14.2);
\node[rval, text=lemon, anchor=base] at (94.1,13) {20};
\node[rtick, anchor=base west] at (127.5,13.1) {26 solved};
\node[rtick, anchor=base east] at (41.5,18.4) {with listing};
\fill[taupe!55] (43,15.3) rectangle (97.6,19.5);
\node[rval, text=lemon, anchor=base] at (70.3,18.3) {79};
\draw[white, line width=0.5pt] (97.6,15.3) -- (97.6,19.5);
\fill[coral] (97.6,15.3) rectangle (106.6,19.5);
\node[rval, text=coffee, anchor=base] at (102.1,18.3) {13};
\draw[white, line width=0.5pt] (106.6,15.3) -- (106.6,19.5);
\fill[cinnamon!62!lemon] (106.6,15.3) rectangle (119.8,19.5);
\node[rval, text=coffee, anchor=base] at (113.2,18.3) {19};
\draw[white, line width=0.5pt] (119.8,15.3) -- (119.8,19.5);
\fill[coffee] (119.8,15.3) rectangle (126,19.5);
\node[rval, text=lemon, anchor=base] at (122.9,18.3) {9};
\node[rtick, anchor=base west] at (127.5,18.4) {120 solved};
\node[rttl, anchor=base west] at (0,26.2) {Qwen3-235B};
\node[rtick, anchor=base east] at (41.5,26.1) {bits only};
\fill[taupe!55] (43,23) rectangle (56.8,27.2);
\node[rval, text=lemon, anchor=base] at (49.9,26) {1};
\draw[white, line width=0.5pt] (56.8,23) -- (56.8,27.2);
\fill[coffee] (56.8,23) rectangle (126,27.2);
\node[rval, text=lemon, anchor=base] at (91.4,26) {5};
\node[rtick, anchor=base west] at (127.5,26.1) {6 solved};
\node[rtick, anchor=base east] at (41.5,31.4) {with listing};
\fill[taupe!55] (43,28.3) rectangle (113.7,32.5);
\node[rval, text=lemon, anchor=base] at (78.4,31.3) {46};
\draw[white, line width=0.5pt] (113.7,28.3) -- (113.7,32.5);
\fill[coral] (113.7,28.3) rectangle (118.3,32.5);
\draw[white, line width=0.5pt] (118.3,28.3) -- (118.3,32.5);
\fill[cinnamon!62!lemon] (118.3,28.3) rectangle (124.5,32.5);
\node[rval, text=coffee, anchor=base] at (121.4,31.3) {4};
\draw[white, line width=0.5pt] (124.5,28.3) -- (124.5,32.5);
\fill[coffee] (124.5,28.3) rectangle (126,32.5);
\node[rtick, anchor=base west] at (127.5,31.4) {54 solved};
\node[rttl, anchor=base west] at (0,39.2) {Nemotron-3 Super};
\node[rtick, anchor=base east] at (41.5,39.1) {bits only};
\fill[coffee] (43,36) rectangle (126,40.2);
\node[rval, text=lemon, anchor=base] at (84.5,39) {3};
\node[rtick, anchor=base west] at (127.5,39.1) {3 solved};
\node[rtick, anchor=base east] at (41.5,44.4) {with listing};
\fill[taupe!55] (43,41.3) rectangle (102.6,45.5);
\node[rval, text=lemon, anchor=base] at (72.8,44.3) {28};
\draw[white, line width=0.5pt] (102.6,41.3) -- (102.6,45.5);
\fill[coral] (102.6,41.3) rectangle (106.8,45.5);
\draw[white, line width=0.5pt] (106.8,41.3) -- (106.8,45.5);
\fill[cinnamon!62!lemon] (106.8,41.3) rectangle (111.1,45.5);
\draw[white, line width=0.5pt] (111.1,41.3) -- (111.1,45.5);
\fill[coffee] (111.1,41.3) rectangle (126,45.5);
\node[rval, text=lemon, anchor=base] at (118.6,44.3) {7};
\node[rtick, anchor=base west] at (127.5,44.4) {39 solved};
\node[rttl, anchor=base west] at (0,52.2) {MiniMax-M3};
\node[rtick, anchor=base east] at (41.5,52.1) {bits only};
\fill[taupe!55] (43,49) rectangle (98.3,53.2);
\node[rval, text=lemon, anchor=base] at (70.7,52) {4};
\draw[white, line width=0.5pt] (98.3,49) -- (98.3,53.2);
\fill[cinnamon!62!lemon] (98.3,49) rectangle (112.2,53.2);
\node[rval, text=coffee, anchor=base] at (105.3,52) {1};
\draw[white, line width=0.5pt] (112.2,49) -- (112.2,53.2);
\fill[coffee] (112.2,49) rectangle (126,53.2);
\node[rval, text=lemon, anchor=base] at (119.1,52) {1};
\node[rtick, anchor=base west] at (127.5,52.1) {6 solved};
\node[rtick, anchor=base east] at (41.5,57.4) {with listing};
\fill[taupe!55] (43,54.3) rectangle (103.3,58.5);
\node[rval, text=lemon, anchor=base] at (73.1,57.3) {69};
\draw[white, line width=0.5pt] (103.3,54.3) -- (103.3,58.5);
\fill[coral] (103.3,54.3) rectangle (118.1,58.5);
\node[rval, text=coffee, anchor=base] at (110.7,57.3) {17};
\draw[white, line width=0.5pt] (118.1,54.3) -- (118.1,58.5);
\fill[cinnamon!62!lemon] (118.1,54.3) rectangle (124.3,58.5);
\node[rval, text=coffee, anchor=base] at (121.2,57.3) {7};
\draw[white, line width=0.5pt] (124.3,54.3) -- (124.3,58.5);
\fill[coffee] (124.3,54.3) rectangle (126,58.5);
\node[rtick, anchor=base west] at (127.5,57.4) {95 solved};
\node[rttl, anchor=base west] at (0,65.2) {DeepSeek-V4.1-Flash};
\node[rtick, anchor=base east] at (41.5,65.1) {bits only};
\fill[taupe!55] (43,62) rectangle (73,66.2);
\node[rval, text=lemon, anchor=base] at (58,65) {60};
\draw[white, line width=0.5pt] (73,62) -- (73,66.2);
\fill[coral] (73,62) rectangle (107,66.2);
\node[rval, text=coffee, anchor=base] at (90,65) {68};
\draw[white, line width=0.5pt] (107,62) -- (107,66.2);
\fill[cinnamon!62!lemon] (107,62) rectangle (117.5,66.2);
\node[rval, text=coffee, anchor=base] at (112.2,65) {21};
\draw[white, line width=0.5pt] (117.5,62) -- (117.5,66.2);
\fill[coffee] (117.5,62) rectangle (126,66.2);
\node[rval, text=lemon, anchor=base] at (121.8,65) {17};
\node[rtick, anchor=base west] at (127.5,65.1) {166 solved};
\node[rtick, anchor=base east] at (41.5,70.4) {with listing};
\fill[taupe!55] (43,67.3) rectangle (75.9,71.5);
\node[rval, text=lemon, anchor=base] at (59.5,70.3) {165};
\draw[white, line width=0.5pt] (75.9,67.3) -- (75.9,71.5);
\fill[coral] (75.9,67.3) rectangle (110,71.5);
\node[rval, text=coffee, anchor=base] at (93,70.3) {171};
\draw[white, line width=0.5pt] (110,67.3) -- (110,71.5);
\fill[cinnamon!62!lemon] (110,67.3) rectangle (125.6,71.5);
\node[rval, text=coffee, anchor=base] at (117.8,70.3) {78};
\draw[white, line width=0.5pt] (125.6,67.3) -- (125.6,71.5);
\fill[coffee] (125.6,67.3) rectangle (126,71.5);
\node[rtick, anchor=base west] at (127.5,70.4) {416 solved};
\node[rttl, anchor=base west] at (0,78.2) {Claude Opus 5.5};
\node[rtick, anchor=base east] at (41.5,78.1) {bits only};
\fill[taupe!55] (43,75) rectangle (78.1,79.2);
\node[rval, text=lemon, anchor=base] at (60.6,78) {176};
\draw[white, line width=0.5pt] (78.1,75) -- (78.1,79.2);
\fill[coral] (78.1,75) rectangle (119.4,79.2);
\node[rval, text=coffee, anchor=base] at (98.8,78) {207};
\draw[white, line width=0.5pt] (119.4,75) -- (119.4,79.2);
\fill[cinnamon!62!lemon] (119.4,75) rectangle (126,79.2);
\node[rval, text=coffee, anchor=base] at (122.7,78) {33};
\node[rtick, anchor=base west] at (127.5,78.1) {416 solved};
\node[rtick, anchor=base east] at (41.5,83.4) {with listing};
\fill[taupe!55] (43,80.3) rectangle (76.7,84.5);
\node[rval, text=lemon, anchor=base] at (59.9,83.3) {172};
\draw[white, line width=0.5pt] (76.7,80.3) -- (76.7,84.5);
\fill[coral] (76.7,80.3) rectangle (117.8,84.5);
\node[rval, text=coffee, anchor=base] at (97.3,83.3) {209};
\draw[white, line width=0.5pt] (117.8,80.3) -- (117.8,84.5);
\fill[cinnamon!62!lemon] (117.8,80.3) rectangle (126,84.5);
\node[rval, text=coffee, anchor=base] at (121.9,83.3) {42};
\node[rtick, anchor=base west] at (127.5,83.4) {423 solved};
\node[rtick, anchor=base east] at (126,89) {numbers inside the bars are runs; each bar is the solved runs of that row};
\end{tikzpicture}
  }
  \caption{How the solved runs changed the file, on the tasks that edit instructions. \emph{The compiler's bytes}: the run changes exactly the bytes that differ in the compiled new version. \emph{Other bytes}: it changes instructions, but a different set of bytes, fewer or not. \emph{Data only}: it changes no instruction and reaches the new behaviour by editing data.}
  \Description{Two stacked bars per model, one for the bits only and one with the disassembler listing. Each bar splits the solved runs into four groups: the compiler's bytes, other bytes that are fewer, other bytes that are as many or more, and data only.}
  \label{fig:routes}
\end{figure}

\paragraph{With the bits only, three models solve mostly the data families}
Supplementary Figure~S1 shows the runs solved in each family. With the bits only, gpt-oss-120b, Qwen3-235B and Nemotron-3 Super solve almost nothing outside the three data families (V3, V5, D1) and family D7, whose change request negates a test. In family D7 the reference solution changes one instruction byte, but the solved runs reach the new behaviour through the data. Figure~\ref{fig:behaviours}b draws one of these runs, where the edit swaps the two texts that the program prints. The same route explains most of what gpt-oss-120b solves on instruction tasks with the bits only: Figure~\ref{fig:routes} shows that 20 of its 26 solved runs change no instruction.

\paragraph{With the listing, four models rarely solve insert logic}
With the listing, the four models solve part of the value runs and the decision runs on instructions, but at most 6.3 percent of the 175 insert logic runs, the last column of Figure~\ref{fig:kinds}. DeepSeek-V4.1-Flash and Claude Opus 5.5 solve nearly all runs of every kind with the listing. For DeepSeek-V4.1-Flash with the bits only, insert logic is not the weakest kind: it solves these runs more often than value changes to instructions. Supplementary Table~S6 makes the same comparison with the family as the unit: the insert logic families are solved less often than the other instruction families by the four models, and not by DeepSeek-V4.1-Flash or Claude Opus 5.5.

\paragraph{With the listing, four models solve fewer runs as the reference solution changes more bytes}
Supplementary Figure~S2 shows the runs solved by the number of bytes that the reference solution changes. With the listing, gpt-oss-120b solves most runs at 1 byte and few at 17 or more bytes. Qwen3-235B, Nemotron-3 Super and MiniMax-M3 fall in the same way. DeepSeek-V4.1-Flash and Claude Opus 5.5 stay high at every size.

\paragraph{The size of the edit and the kind of change overlap}
In Table~\ref{tab:families}, the insert logic families also change the most bytes, so Supplementary Figure~S2 and the insert logic result above are not independent. A comparison inside one kind separates them. Among the value families of Supplementary Figure~S1, with the listing, V1 and V4 (2 to 7 bytes) are solved far more often than V2 and V6 (up to 29 bytes) by the four models. Inside the value kind, the families with larger reference solutions are solved less often.

\subsection{RQ3: Where the run fails}
\label{sec:rq3}

RQ2 shows which kinds of change are solved least; RQ3 asks at which step the unsolved runs fail.

\paragraph{Most unsolved runs fail before the tests}
Figure~\ref{fig:outcomes-results} shows the outcome of every run. Of the 3,519 unsolved runs, 548 reach the tests and fail them; the rest are cut, not applicable or give no edit. Cut is the most common failure of Nemotron-3 Super, MiniMax-M3 and DeepSeek-V4.1-Flash with the bits only. Not applicable is the most common failure of gpt-oss-120b and Qwen3-235B.

\paragraph{For two models, the failing step moves from locating to writing}
gpt-oss-120b and Qwen3-235B have the most test failures with the bits only; the other models have fewer than ten each. With the bits only, wrong place is 90 of the 103 test failures of gpt-oss-120b. With the listing, wrong change is the most common test failure of gpt-oss-120b, Qwen3-235B, MiniMax-M3 and Nemotron-3 Super: the edit changes the right region with a wrong new instruction.

\paragraph{The old behaviour often breaks, but rarely alone}
The outcome broke the rest occurs in 37 of the 6,000 runs, against 219 for wrong place and 292 for wrong change. The checks assign a run to wrong place or wrong change before they can assign it to broke the rest, so this count is the number of runs whose only failure is in the old behaviour. Supplementary Table~S7 counts the tests of the old behaviour on their own: they fail in more than half of the runs that reach the tests. Preserving the rest is therefore rarely the only step that fails, and often one of several.

\subsection{Further observations}
\label{sec:further}

Three observations fall outside the three questions. Two concern the form of the edits, and one the output limit.

\paragraph{Byte-order slips}
Many edits that cannot be applied copy an instruction in the byte order of the listing. Figure~\ref{fig:behaviours}a draws one such slip, and Table~\ref{tab:byteorder} counts them for each model. With the listing, a quarter to nearly two thirds of the not applicable runs of the four models are byte-order slips; DeepSeek-V4.1-Flash and Claude Opus 5.5 have none. With the bits only, the six models together produce 14 slips, against 228 with the listing. Reversing the bytes of the edit would solve 65 of the 242 slips. For Qwen3-235B, the reversed slips would add about a fifth to its solved runs with the listing. Even with this correction, the four models stay far below DeepSeek-V4.1-Flash and Claude Opus 5.5.

\begin{table}[t]
\caption{Byte-order slips. A disassembler prints each 4-byte instruction as one number, most significant byte first, while the file stores the same instruction least significant byte first. \emph{Slips} counts the edits that cannot be applied as written but whose sequence to replace is found exactly once in the file after reversing the bytes of each 4-byte word. \emph{Solved if reversed} re-scores those edits with both sequences reversed.}
\label{tab:byteorder}
\footnotesize
\setlength{\tabcolsep}{4pt}
\begin{tabular}{lrrrrr}
\toprule
 & \multicolumn{2}{c}{Bits only} & \multicolumn{2}{c}{With listing} & \\
\cmidrule(lr){2-3}\cmidrule(lr){4-5}
Model & not applicable & slips & not applicable & slips & Solved if reversed \\
\midrule
gpt-oss-120b & 239 & 4 & 187 & 47 & 7 \\
Qwen3-235B & 255 & 9 & 205 & 106 & 23 \\
Nemotron-3 Super & 12 & 0 & 40 & 23 & 11 \\
MiniMax-M3 & 15 & 1 & 84 & 52 & 24 \\
DeepSeek-V4.1-Flash & 5 & 0 & 4 & 0 & 0 \\
Claude Opus 5.5 & 1 & 0 & 0 & 0 & 0 \\
\bottomrule
\end{tabular}
\end{table}

\paragraph{Edits that pass the tests are not always the compiler's edit}
Figure~\ref{fig:routes} classifies the solved runs on instruction tasks by the bytes they change. With the listing, most solved runs of the four models change exactly the bytes of the compiler. DeepSeek-V4.1-Flash often does not: in 171 of its 416 solved runs with the listing it changes fewer bytes than the compiler does. Claude Opus 5.5 does so in about half of its solved runs: 209 of 423 with the listing and 207 of 416 with the bits only. Figure~\ref{fig:behaviours}c shows one such edit, which removes the instruction that sets the frame pointer and puts a test that jumps to code already in the program.

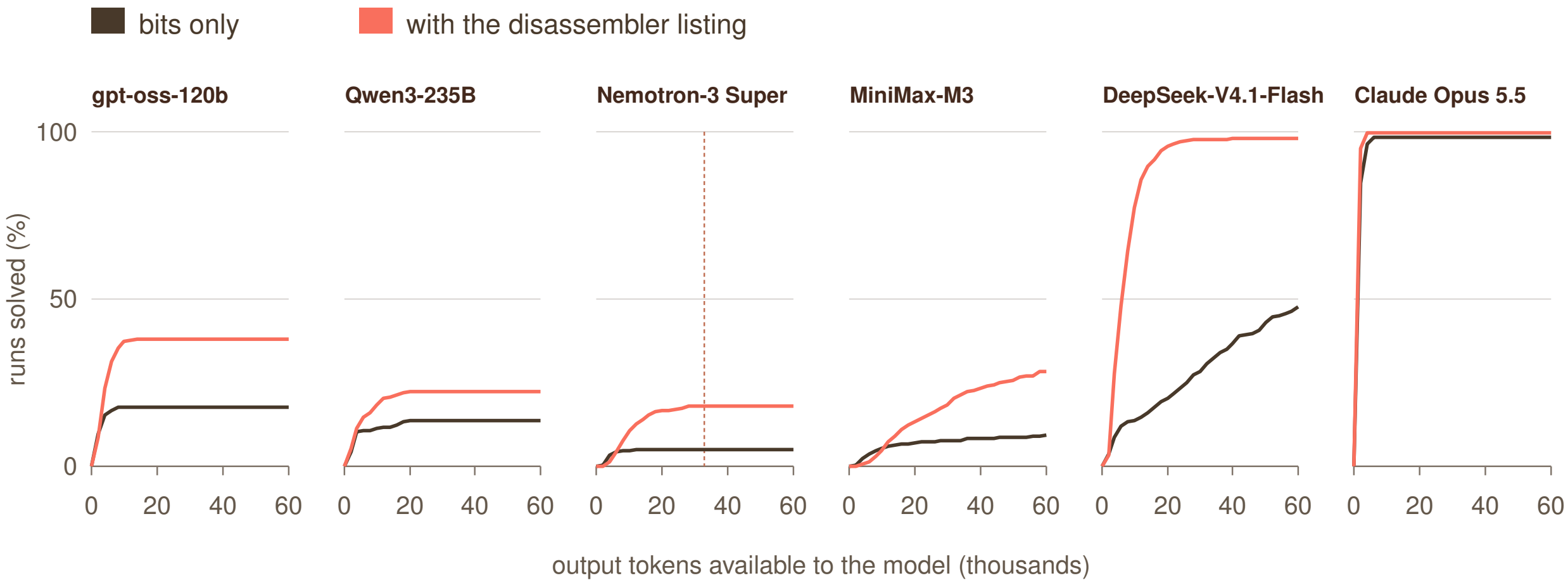
\begin{figure}[t]
  \centering
  \resizebox{\linewidth}{!}{
\begin{tikzpicture}[x=1mm, y=-1mm]
\useasboundingbox (0,0) rectangle (139,53.5);
\fill[taupe] (8,0.9000000000000004) rectangle (11,3.2);
\node[rlab, anchor=base west] at (12.2,3.2) {bits only};
\fill[coral] (32,0.9000000000000004) rectangle (35,3.2);
\node[rlab, anchor=base west] at (36.2,3.2) {with the disassembler listing};
\node[rttl, anchor=base west, scale=0.7] at (8,9.4) {gpt-oss-120b};
\draw[rgrid] (8,42) -- (25.7,42);
\node[rtick, anchor=east] at (6.8,42) {0};
\draw[rgrid] (8,27) -- (25.7,27);
\node[rtick, anchor=east] at (6.8,27) {50};
\draw[rgrid] (8,12) -- (25.7,12);
\node[rtick, anchor=east] at (6.8,12) {100};
\draw[raxis] (8,42) -- (25.7,42);
\draw[raxis] (8,42) -- (8,42.8);
\node[rtick, anchor=base] at (8,46.3) {0};
\draw[raxis] (13.9,42) -- (13.9,42.8);
\node[rtick, anchor=base] at (13.9,46.3) {20};
\draw[raxis] (19.8,42) -- (19.8,42.8);
\node[rtick, anchor=base] at (19.8,46.3) {40};
\draw[raxis] (25.7,42) -- (25.7,42.8);
\node[rtick, anchor=base] at (25.7,46.3) {60};
\draw[taupe, line width=0.9pt, line join=round] (8,42) -- (8.6,39.1) -- (9.2,37.4) -- (9.8,37) -- (10.4,36.7) -- (10.9,36.7) -- (11.5,36.7) -- (12.1,36.7) -- (12.7,36.7) -- (13.3,36.7) -- (13.9,36.7) -- (14.5,36.7) -- (15.1,36.7) -- (15.7,36.7) -- (16.2,36.7) -- (16.8,36.7) -- (17.4,36.7) -- (18,36.7) -- (18.6,36.7) -- (19.2,36.7) -- (19.8,36.7) -- (20.4,36.7) -- (21,36.7) -- (21.5,36.7) -- (22.1,36.7) -- (22.7,36.7) -- (23.3,36.7) -- (23.9,36.7) -- (24.5,36.7) -- (25.1,36.7) -- (25.7,36.7);
\draw[coral, line width=0.9pt, line join=round] (8,42) -- (8.6,39.4) -- (9.2,35) -- (9.8,32.6) -- (10.4,31.4) -- (10.9,30.8) -- (11.5,30.7) -- (12.1,30.6) -- (12.7,30.6) -- (13.3,30.6) -- (13.9,30.6) -- (14.5,30.6) -- (15.1,30.6) -- (15.7,30.6) -- (16.2,30.6) -- (16.8,30.6) -- (17.4,30.6) -- (18,30.6) -- (18.6,30.6) -- (19.2,30.6) -- (19.8,30.6) -- (20.4,30.6) -- (21,30.6) -- (21.5,30.6) -- (22.1,30.6) -- (22.7,30.6) -- (23.3,30.6) -- (23.9,30.6) -- (24.5,30.6) -- (25.1,30.6) -- (25.7,30.6);
\node[rttl, anchor=base west, scale=0.7] at (30.7,9.4) {Qwen3-235B};
\draw[rgrid] (30.7,42) -- (48.3,42);
\draw[rgrid] (30.7,27) -- (48.3,27);
\draw[rgrid] (30.7,12) -- (48.3,12);
\draw[raxis] (30.7,42) -- (48.3,42);
\draw[raxis] (30.7,42) -- (30.7,42.8);
\node[rtick, anchor=base] at (30.7,46.3) {0};
\draw[raxis] (36.6,42) -- (36.6,42.8);
\node[rtick, anchor=base] at (36.6,46.3) {20};
\draw[raxis] (42.4,42) -- (42.4,42.8);
\node[rtick, anchor=base] at (42.4,46.3) {40};
\draw[raxis] (48.3,42) -- (48.3,42.8);
\node[rtick, anchor=base] at (48.3,46.3) {60};
\draw[taupe, line width=0.9pt, line join=round] (30.7,42) -- (31.3,40.7) -- (31.8,38.9) -- (32.4,38.8) -- (33,38.8) -- (33.6,38.6) -- (34.2,38.5) -- (34.8,38.5) -- (35.4,38.3) -- (36,38) -- (36.6,37.9) -- (37.1,37.9) -- (37.7,37.9) -- (38.3,37.9) -- (38.9,37.9) -- (39.5,37.9) -- (40.1,37.9) -- (40.7,37.9) -- (41.3,37.9) -- (41.9,37.9) -- (42.4,37.9) -- (43,37.9) -- (43.6,37.9) -- (44.2,37.9) -- (44.8,37.9) -- (45.4,37.9) -- (46,37.9) -- (46.6,37.9) -- (47.2,37.9) -- (47.7,37.9) -- (48.3,37.9);
\draw[coral, line width=0.9pt, line join=round] (30.7,42) -- (31.3,40.4) -- (31.8,38.6) -- (32.4,37.6) -- (33,37.2) -- (33.6,36.5) -- (34.2,35.9) -- (34.8,35.8) -- (35.4,35.6) -- (36,35.4) -- (36.6,35.3) -- (37.1,35.3) -- (37.7,35.3) -- (38.3,35.3) -- (38.9,35.3) -- (39.5,35.3) -- (40.1,35.3) -- (40.7,35.3) -- (41.3,35.3) -- (41.9,35.3) -- (42.4,35.3) -- (43,35.3) -- (43.6,35.3) -- (44.2,35.3) -- (44.8,35.3) -- (45.4,35.3) -- (46,35.3) -- (46.6,35.3) -- (47.2,35.3) -- (47.7,35.3) -- (48.3,35.3);
\node[rttl, anchor=base west, scale=0.7] at (53.3,9.4) {Nemotron-3 Super};
\draw[rgrid] (53.3,42) -- (71,42);
\draw[rgrid] (53.3,27) -- (71,27);
\draw[rgrid] (53.3,12) -- (71,12);
\draw[raxis] (53.3,42) -- (71,42);
\draw[raxis] (53.3,42) -- (53.3,42.8);
\node[rtick, anchor=base] at (53.3,46.3) {0};
\draw[raxis] (59.2,42) -- (59.2,42.8);
\node[rtick, anchor=base] at (59.2,46.3) {20};
\draw[raxis] (65.1,42) -- (65.1,42.8);
\node[rtick, anchor=base] at (65.1,46.3) {40};
\draw[raxis] (71,42) -- (71,42.8);
\node[rtick, anchor=base] at (71,46.3) {60};
\draw[cinnamon, line width=0.4pt, dash pattern=on 1pt off 0.8pt] (63,12) -- (63,42);
\draw[taupe, line width=0.9pt, line join=round] (53.3,42) -- (53.9,41.9) -- (54.5,41) -- (55.1,40.7) -- (55.7,40.6) -- (56.3,40.6) -- (56.9,40.5) -- (57.5,40.5) -- (58,40.5) -- (58.6,40.5) -- (59.2,40.5) -- (59.8,40.5) -- (60.4,40.5) -- (61,40.5) -- (61.6,40.5) -- (62.2,40.5) -- (62.8,40.5) -- (63.3,40.5) -- (63.9,40.5) -- (64.5,40.5) -- (65.1,40.5) -- (65.7,40.5) -- (66.3,40.5) -- (66.9,40.5) -- (67.5,40.5) -- (68.1,40.5) -- (68.6,40.5) -- (69.2,40.5) -- (69.8,40.5) -- (70.4,40.5) -- (71,40.5);
\draw[coral, line width=0.9pt, line join=round] (53.3,42) -- (53.9,42) -- (54.5,41.6) -- (55.1,40.7) -- (55.7,39.7) -- (56.3,38.8) -- (56.9,38.2) -- (57.5,37.8) -- (58,37.4) -- (58.6,37.1) -- (59.2,37) -- (59.8,37) -- (60.4,36.9) -- (61,36.8) -- (61.6,36.6) -- (62.2,36.6) -- (62.8,36.6) -- (63.3,36.6) -- (63.9,36.6) -- (64.5,36.6) -- (65.1,36.6) -- (65.7,36.6) -- (66.3,36.6) -- (66.9,36.6) -- (67.5,36.6) -- (68.1,36.6) -- (68.6,36.6) -- (69.2,36.6) -- (69.8,36.6) -- (70.4,36.6) -- (71,36.6);
\node[rttl, anchor=base west, scale=0.7] at (76,9.4) {MiniMax-M3};
\draw[rgrid] (76,42) -- (93.7,42);
\draw[rgrid] (76,27) -- (93.7,27);
\draw[rgrid] (76,12) -- (93.7,12);
\draw[raxis] (76,42) -- (93.7,42);
\draw[raxis] (76,42) -- (76,42.8);
\node[rtick, anchor=base] at (76,46.3) {0};
\draw[raxis] (81.9,42) -- (81.9,42.8);
\node[rtick, anchor=base] at (81.9,46.3) {20};
\draw[raxis] (87.8,42) -- (87.8,42.8);
\node[rtick, anchor=base] at (87.8,46.3) {40};
\draw[raxis] (93.7,42) -- (93.7,42.8);
\node[rtick, anchor=base] at (93.7,46.3) {60};
\draw[taupe, line width=0.9pt, line join=round] (76,42) -- (76.6,41.9) -- (77.2,41.3) -- (77.8,40.9) -- (78.4,40.6) -- (78.9,40.4) -- (79.5,40.2) -- (80.1,40.1) -- (80.7,40) -- (81.3,40) -- (81.9,39.9) -- (82.5,39.8) -- (83.1,39.8) -- (83.7,39.8) -- (84.2,39.7) -- (84.8,39.7) -- (85.4,39.7) -- (86,39.7) -- (86.6,39.5) -- (87.2,39.5) -- (87.8,39.5) -- (88.4,39.5) -- (89,39.5) -- (89.5,39.4) -- (90.1,39.4) -- (90.7,39.4) -- (91.3,39.4) -- (91.9,39.4) -- (92.5,39.3) -- (93.1,39.3) -- (93.7,39.2);
\draw[coral, line width=0.9pt, line join=round] (76,42) -- (76.6,42) -- (77.2,41.8) -- (77.8,41.6) -- (78.4,41.1) -- (78.9,40.6) -- (79.5,39.8) -- (80.1,39.3) -- (80.7,38.7) -- (81.3,38.3) -- (81.9,38) -- (82.5,37.7) -- (83.1,37.4) -- (83.7,37.1) -- (84.2,36.8) -- (84.8,36.5) -- (85.4,35.9) -- (86,35.6) -- (86.6,35.3) -- (87.2,35.2) -- (87.8,35) -- (88.4,34.8) -- (89,34.7) -- (89.5,34.5) -- (90.1,34.4) -- (90.7,34.3) -- (91.3,34) -- (91.9,33.9) -- (92.5,33.9) -- (93.1,33.5) -- (93.7,33.5);
\node[rttl, anchor=base west, scale=0.7] at (98.7,9.4) {DeepSeek-V4.1-Flash};
\draw[rgrid] (98.7,42) -- (116.3,42);
\draw[rgrid] (98.7,27) -- (116.3,27);
\draw[rgrid] (98.7,12) -- (116.3,12);
\draw[raxis] (98.7,42) -- (116.3,42);
\draw[raxis] (98.7,42) -- (98.7,42.8);
\node[rtick, anchor=base] at (98.7,46.3) {0};
\draw[raxis] (104.6,42) -- (104.6,42.8);
\node[rtick, anchor=base] at (104.6,46.3) {20};
\draw[raxis] (110.4,42) -- (110.4,42.8);
\node[rtick, anchor=base] at (110.4,46.3) {40};
\draw[raxis] (116.3,42) -- (116.3,42.8);
\node[rtick, anchor=base] at (116.3,46.3) {60};
\draw[taupe, line width=0.9pt, line join=round] (98.7,42) -- (99.3,40.9) -- (99.8,39.4) -- (100.4,38.4) -- (101,38) -- (101.6,37.9) -- (102.2,37.6) -- (102.8,37.2) -- (103.4,36.7) -- (104,36.2) -- (104.6,35.9) -- (105.1,35.5) -- (105.7,35) -- (106.3,34.5) -- (106.9,33.8) -- (107.5,33.5) -- (108.1,32.8) -- (108.7,32.3) -- (109.3,31.8) -- (109.9,31.5) -- (110.4,31) -- (111,30.3) -- (111.6,30.2) -- (112.2,30.1) -- (112.8,29.8) -- (113.4,29.1) -- (114,28.6) -- (114.6,28.5) -- (115.2,28.3) -- (115.7,28.1) -- (116.3,27.7);
\draw[coral, line width=0.9pt, line join=round] (98.7,42) -- (99.3,41) -- (99.8,33.7) -- (100.4,27.6) -- (101,22.7) -- (101.6,18.8) -- (102.2,16.3) -- (102.8,15.1) -- (103.4,14.5) -- (104,13.7) -- (104.6,13.3) -- (105.1,13.1) -- (105.7,12.9) -- (106.3,12.8) -- (106.9,12.7) -- (107.5,12.7) -- (108.1,12.7) -- (108.7,12.7) -- (109.3,12.7) -- (109.9,12.7) -- (110.4,12.6) -- (111,12.6) -- (111.6,12.6) -- (112.2,12.6) -- (112.8,12.6) -- (113.4,12.6) -- (114,12.6) -- (114.6,12.6) -- (115.2,12.6) -- (115.7,12.6) -- (116.3,12.6);
\node[rttl, anchor=base west, scale=0.7] at (121.3,9.4) {Claude Opus 5.5};
\draw[rgrid] (121.3,42) -- (139,42);
\draw[rgrid] (121.3,27) -- (139,27);
\draw[rgrid] (121.3,12) -- (139,12);
\draw[raxis] (121.3,42) -- (139,42);
\draw[raxis] (121.3,42) -- (121.3,42.8);
\node[rtick, anchor=base] at (121.3,46.3) {0};
\draw[raxis] (127.2,42) -- (127.2,42.8);
\node[rtick, anchor=base] at (127.2,46.3) {20};
\draw[raxis] (133.1,42) -- (133.1,42.8);
\node[rtick, anchor=base] at (133.1,46.3) {40};
\draw[raxis] (139,42) -- (139,42.8);
\node[rtick, anchor=base] at (139,46.3) {60};
\draw[taupe, line width=0.9pt, line join=round] (121.3,42) -- (121.9,16.6) -- (122.5,13.1) -- (123.1,12.5) -- (123.7,12.5) -- (124.3,12.5) -- (124.9,12.5) -- (125.5,12.5) -- (126,12.5) -- (126.6,12.5) -- (127.2,12.5) -- (127.8,12.5) -- (128.4,12.5) -- (129,12.5) -- (129.6,12.5) -- (130.2,12.5) -- (130.8,12.5) -- (131.3,12.5) -- (131.9,12.5) -- (132.5,12.5) -- (133.1,12.5) -- (133.7,12.5) -- (134.3,12.5) -- (134.9,12.5) -- (135.5,12.5) -- (136.1,12.5) -- (136.6,12.5) -- (137.2,12.5) -- (137.8,12.5) -- (138.4,12.5) -- (139,12.5);
\draw[coral, line width=0.9pt, line join=round] (121.3,42) -- (121.9,13.5) -- (122.5,12.1) -- (123.1,12.1) -- (123.7,12.1) -- (124.3,12.1) -- (124.9,12.1) -- (125.5,12.1) -- (126,12.1) -- (126.6,12.1) -- (127.2,12.1) -- (127.8,12.1) -- (128.4,12.1) -- (129,12.1) -- (129.6,12.1) -- (130.2,12.1) -- (130.8,12.1) -- (131.3,12.1) -- (131.9,12.1) -- (132.5,12.1) -- (133.1,12.1) -- (133.7,12.1) -- (134.3,12.1) -- (134.9,12.1) -- (135.5,12.1) -- (136.1,12.1) -- (136.6,12.1) -- (137.2,12.1) -- (137.8,12.1) -- (138.4,12.1) -- (139,12.1);
\node[rtick, anchor=base, rotate=90] at (2.0,27.0) {runs solved (\%)};
\node[rtick, anchor=base] at (73.5,51.6) {output tokens available to the model (thousands)};
\end{tikzpicture}
  }
  \caption{Runs solved as the output budget grows: for each number of output tokens, the share of runs whose answer was complete and correct within that budget. Each run had one attempt and up to 60,000 tokens, so a run counts as solved at a smaller budget when its complete answer fits within it. A dashed line marks the point where the provider stops a model before 60,000 tokens.}
  \Description{One panel per model. Each panel has two rising curves over output tokens from 0 to 60,000: the share of runs solved with the bits only and with the disassembler listing.}
  \label{fig:budget}
\end{figure}

\paragraph{For DeepSeek-V4.1-Flash and MiniMax-M3, the runs solved depend on the output limit}
Figure~\ref{fig:budget} shows the runs solved as the output budget grows. DeepSeek-V4.1-Flash with the bits only solves 102 runs within 20,000 tokens and 238 within 60,000, so its result depends on the limit; with the listing it solves nearly all runs within 20,000. MiniMax-M3 with the listing also gains between 20,000 and 60,000 tokens. Every cut run of Nemotron-3 Super ends at 32,768 tokens, the limit that its provider sets and that Section~S1 of the supplementary material records. On that same limit MiniMax-M3 solves more runs than Nemotron-3 Super, with the bits only and with the listing. Claude Opus 5.5 finishes every run within 5,561 output tokens, so its result does not depend on the limit.

\subsection{One model with little decoding bottleneck}
\label{sec:frontier}

In Table~\ref{tab:models}, Claude Opus 5.5 stands apart: it solves 491 of 500 runs with the bits only and 498 of 500 with the listing, its output is the shortest of all models, and none of its runs is cut. Only 11 of its 1,000 runs are unsolved. They share a shape, and the shape shows where the difficulty lies once decoding is not the limit.

\begin{figure}[t]
  \centering
  \resizebox{\linewidth}{!}{
\begin{tikzpicture}[x=1mm, y=-1mm,
  bt/.style={anchor=base west, inner sep=0, text=coffee, font=\fontfamily{phv}\fontsize{8.8}{10.5}\selectfont\bfseries},
  bd/.style={anchor=base west, inner sep=0, text=taupe, font=\fontfamily{phv}\fontsize{7}{8.8}\selectfont},
  bi/.style={anchor=base west, inner sep=0, text=taupe!80, font=\fontfamily{phv}\fontsize{6.5}{8}\selectfont\itshape},
  bbits/.style={anchor=base, inner sep=0, text=coffee, font=\ttfamily\fontsize{7}{7.5}\selectfont},
  bs/.style={anchor=base, inner sep=0, text=coffee, font=\ttfamily\fontsize{5.8}{7.5}\selectfont},
  bnm/.style={anchor=base west, inner sep=0, text=coffee, font=\fontfamily{phv}\fontsize{8.2}{10}\selectfont\bfseries},
  bchip/.style={draw=taupe!75, line width=0.45pt, rounded corners=1.2mm, fill=white},
  bcell/.style={draw=taupe!45, line width=0.3pt, rounded corners=0.5mm, fill=white},
  brefhot/.style={draw=coral, line width=0.6pt, rounded corners=0.5mm, fill=white},
  bhot/.style={draw=coral, line width=0.3pt, rounded corners=0.5mm, fill=coral}]
\useasboundingbox (0,0) rectangle (139,33.4);
\fill[lemon, rounded corners=1.2mm] (0,0) rectangle (139,33.4);
\node[bt] at (5.6,7.3) {One bit short};
\node[bd] at (5.6,13) {The new test is in place; the last};
\node[bd] at (5.6,16.4) {instruction keeps the signed};
\node[bd] at (5.6,19.8) {condition of the original.};
\node[bi] at (5.6,24.8) {Opus 5.5, bits only:};
\node[bi] at (5.6,28.6) {8 of its 11 failures in 1,000 runs};
\node[bi] at (46,4.2) {the compiler};
\node[bi] at (75,4.2) {Opus 5.5};
\draw[bcell] (46,5.6) rectangle (69.5,10.0); \node[bs] at (57.75,8.7) {sub x0, x0, \#0x10};
\draw[bcell] (75,5.6) rectangle (98.5,10.0); \node[bs] at (86.75,8.7) {sub x0, x0, \#0x10};
\draw[bcell] (46,11.0) rectangle (69.5,15.4); \node[bs] at (57.75,14.1) {cmp x0, \#0x31};
\draw[bcell] (75,11.0) rectangle (98.5,15.4); \node[bs] at (86.75,14.1) {cmp x0, \#0x31};
\draw[brefhot] (46,16.4) rectangle (69.5,20.8); \node[bs] at (57.75,19.5) {cset w0, hi};
\draw[bhot] (75,16.4) rectangle (98.5,20.8); \node[bs] at (86.75,19.5) {cset w0, gt};
\node[bi] at (46,24.8) {byte 713, in bits};
\node[bbits] at (57.75,28.6) {1\textcolor{coral}{\textbf{0}}01\,0111};
\node[bbits] at (86.75,28.6) {1\textcolor{coral}{\textbf{1}}01\,0111};
\draw[bchip] (102.5,4) rectangle (137.5,30.0);
\node[bnm] at (105.5,10.6) {Wrong change};
\node[bi] at (105.5,15.4) {right place, right instructions;};
\node[bi] at (105.5,18.6) {one bit of byte 713 differs:};
\node[bi] at (105.5,21.8) {gt compares with sign,};
\node[bi] at (105.5,25) {the range test needs hi};
\end{tikzpicture}
  }
  \caption{One bit short. Claude Opus 5.5 with the bits only, task I5\_01 at level A: the edit writes the new test with the compiler's instructions, but keeps the signed condition of the original program in the last one. Byte 713 differs from the compiled new version in one bit. Eight of the 11 failures of this model have this form.}
  \Description{Three instructions written by the compiler beside the three written by the model. The first two are equal; the third reads cset w0, hi for the compiler and cset w0, gt for the model, highlighted. Below, byte 713 in bits for each, with the one differing bit highlighted. On the right, the outcome: wrong change.}
  \label{fig:onebit}
\end{figure}
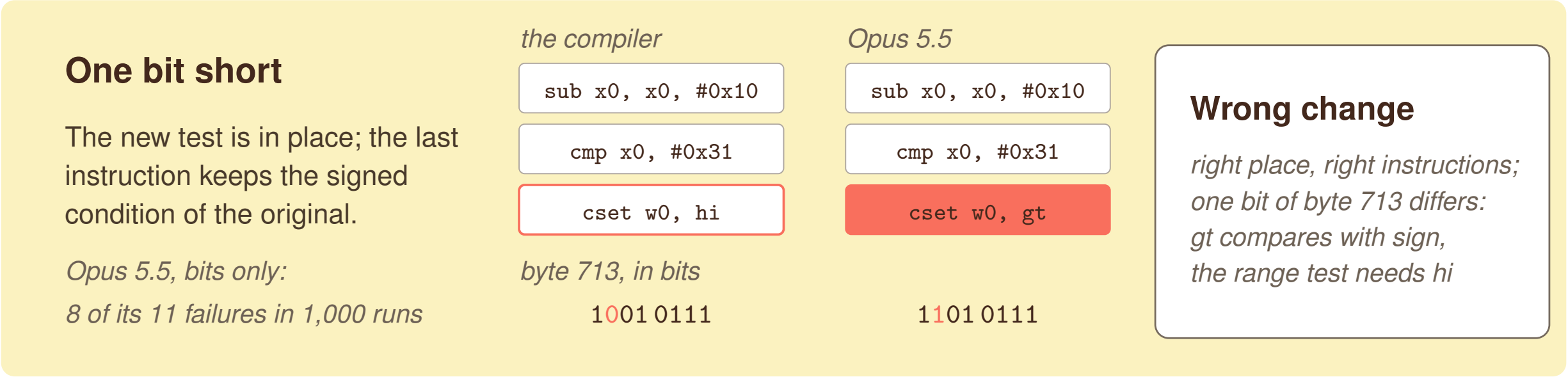

Ten of the 11 runs reach the tests and fail as wrong change: the edit changes the right region, and a test of the new behaviour fails. In each of the ten, the edited code differs from the compiled new version in one bit. Eight of them are in family I5, which asks the program to also reject arguments below a lower bound. All eight are runs with the bits only at levels A and B, the two levels without code; at levels C to E, and with the listing at every level, the same tasks are solved. Figure~\ref{fig:onebit} shows one of the eight. The edit writes the subtraction and the comparison that the compiler writes, but keeps the conditional set of the original program, which compares with sign (\texttt{gt}). The range test needs the comparison without sign (\texttt{hi}), and the two encodings differ in one bit. The other two of the ten are the runs of task D5\_02 at levels C and E with the listing. There, the conditional compare names register \texttt{x0} where the compiler names \texttt{x1}, again one bit.

The eleventh run gives an edit that cannot be applied. Its sequence to replace starts with a run of 12 identical \texttt{nop} instructions, where the file holds 13. Placed where it was meant to go, the same replacement solves the task.

None of the 11 failures is in the wrong place, and none breaks the rest of the program. What fails is a detail of the encoding that the bytes next to the edit do not show: the sign of a comparison, one register field among several, or the length of a repeated run.

Two sets of control runs, outside the 6,000 runs of the study, ask whether the eight failures in family I5 depend on the bit form of the input or on the single attempt. Both use Claude Opus 5.5 with the settings of Section~\ref{sec:models}, at levels A and B, and were run on 24 November 2026. Table~\ref{tab:checks} gives their counts.

The first set repeats, on all 100 tasks, the 200 runs with the bits only at levels A and B, with the executable written as pairs of hexadecimal digits. Apart from the executable, the prompt differs only in the wording that names the form of the input and of the answer, and the edits are written in hexadecimal. This form lies between the two of the study: each byte takes two characters instead of eight, and nothing is decoded into instructions. With it the model solves 194 of the 200 runs, against 191 with the bits and 200 with the listing, and it uses fewer output tokens than with the bits. The six unsolved runs are all in family I5, and all six write the signed condition of Figure~\ref{fig:onebit}. For this model, then, the failure does not depend on the binary digits: it stays when the bytes are given in hexadecimal and disappears when the listing shows the instructions.

The second set calls the model five more times on each of the ten runs of family I5 with the bits only at levels A and B, with the prompt of the study. With the run of the study, each task and level has six calls. Of the 60 calls, 17 are solved. All 43 unsolved calls keep the signed condition, and in 31 of them it is the only bit that differs from the compiled new version. The result of a single task changes between calls: eight of the ten pairs of task and level have solved and unsolved calls. The five repeated rounds of ten calls solve between 1 and 5, so the 2 of 10 of the study lies within their spread.

\begin{table}[t]
\caption{Two sets of control runs for Claude Opus 5.5 at levels A and B, outside the 6,000 runs of the study. (a) The 200 runs on the 100 tasks, by the form of the executable in the prompt; tokens are means over the 200 runs. (b) Family I5 with the bits only: calls solved out of 6 for each task and level, the run of the study and 5 repeated calls of the same prompt.}
\label{tab:checks}
\footnotesize
\setlength{\tabcolsep}{4pt}
\begin{tabular}[t]{lrrrr}
\toprule
 & \multicolumn{2}{c}{Solved of 100} & \multicolumn{2}{c}{Tokens per run} \\
\cmidrule(lr){2-3}\cmidrule(lr){4-5}
(a) Executable given as & level A & level B & output & input \\
\midrule
Bits only & 94 & 97 & 1,442 & 6,661 \\
Hexadecimal only & 96 & 98 & 1,129 & 3,952 \\
Bits and listing & 100 & 100 & 1,059 & 10,602 \\
\bottomrule
\end{tabular}
\hspace{12mm}
\begin{tabular}[t]{lrr}
\toprule
 & \multicolumn{2}{c}{Solved of 6} \\
\cmidrule(lr){2-3}
(b) Task & level A & level B \\
\midrule
I5\_01 & 0 & 1 \\
I5\_02 & 1 & 6 \\
I5\_03 & 1 & 3 \\
I5\_04 & 1 & 2 \\
I5\_05 & 1 & 1 \\
\midrule
All five, of 30 & 4 & 13 \\
\bottomrule
\end{tabular}
\end{table}

\FloatBarrier
\section{Discussion}
\label{sec:discussion}

The results answer the three questions for these executables and models. This section asks what they mean for practice and how they become a method, what they imply for larger executables, and what limits them.

\subsection{Practice and a method for direct editing}
\label{sec:practice}
\label{sec:method}

The results bear first on patching without source. Four elements of a method for direct editing also follow from them: the disassembler listing first, a byte-order check, edits measured against the compiler's bytes, and an output limit set high enough. Each is stated below with the result that suggests it.

\paragraph{Patching without source}
Where the source is missing or the build cannot be reproduced, a model can produce a candidate patch for an executable. In Table~\ref{tab:models}, DeepSeek-V4.1-Flash and Claude Opus 5.5 solve nearly all their runs with the listing and the other four models fewer than two fifths; Claude Opus 5.5 also solves nearly all its runs with the bits only. The choice of model matters. Binary rewriting tools already work in this setting \citep{wenzl2019binaryrewriting}.

\paragraph{The listing comes first}
In the bits-only grid of Figure~\ref{fig:kinds}, four of the six models rarely solve tasks that change instructions, and the tests of Table~\ref{tab:tests} find a gain with the listing for five models. A pipeline for direct editing should therefore run a disassembler before the model is called, as the listing in this study does, and should not rely on the model to decode the bits. Claude Opus 5.5 needs the listing least on these tasks, and the listing adds 7 solved runs. Of the two factors in the prompt, the listing is the only one with a gain for five of the six models. This is the first element.

\paragraph{Decoding has a price in output tokens and money}
By the output tokens of Table~\ref{tab:models}, DeepSeek-V4.1-Flash produces about six times as many per run with the bits only as with the listing. At the list prices of Supplementary Table~S1, a solved run costs 0.11 USD (0.10 EUR) with the bits only and 0.012 USD (0.011 EUR) with the listing. The listing adds input tokens per run and lowers the cost. For the other four models of that provider the cost per solved run also falls with the listing. Claude Opus 5.5, at its own list prices, costs 0.055 USD (0.050 EUR) per solved run with the bits only and 0.064 USD (0.058 EUR) with the listing: its output is short either way, and the listing adds input tokens.

\paragraph{Byte order can be checked by a tool}
A byte-order slip is detected without a model, by the rule that Table~\ref{tab:byteorder} applies: the sequence to replace does not occur in the file as written, but it occurs exactly once after the bytes of each 4-byte word are reversed. A tool can apply this check before it applies an edit, at no cost in model calls. This is the second element. Reversing the slips in our runs would solve 65 runs, the count that Section~\ref{sec:further} computes on the stored runs.

\paragraph{Edits are measured against the compiler's bytes}
Figure~\ref{fig:routes} shows that an edit that passes the tests is often not the one the compiler makes. In about two fifths of the solved runs of DeepSeek-V4.1-Flash on instruction tasks with the listing, and in about half of those of Claude Opus 5.5, the edit changes fewer bytes than the compiler does, and the hidden tests accept it. A direct edit can therefore be a smaller change to the file than a recompilation, which matters when the footprint of a patch is itself a requirement. The study does not test behaviour outside the hidden tests. Other edits leave the instructions as they are and reach the new behaviour through the data, and Figure~\ref{fig:behaviours} draws one edit of each kind. Measuring each edit against the compiler's bytes, as well as against the tests, tells these kinds apart instead of counting them only as solved. This is the third element.

\paragraph{The output limit is a setting}
In Figure~\ref{fig:budget}, the runs solved by DeepSeek-V4.1-Flash and MiniMax-M3 rise between 20,000 and 60,000 output tokens. The limit is therefore the fourth element: it has to be set high enough.

\paragraph{Status of the elements}
These elements come from observing all 100 tasks, and no task was held out. A test of the method needs tasks that the observation did not use, with the first two elements applied by a tool, the third as a measure and the fourth as a setting. 

\subsection{Tasks that may stay hard}
\label{sec:hard}

One model solves nearly every task of this set with the bits only, so a harder evaluation needs tasks beyond it. The 11 failures of Claude Opus 5.5, described in Section~\ref{sec:frontier}, suggest three kinds of task; none of them was tested here. The first asks for a change to a property that the request implies without naming it and that the instruction encodes in one or two bits, such as the sign of a comparison, the width of a register or whether an instruction sets the flags: the eight failures in family I5 keep a signed comparison where the new test needs an unsigned one. The second uses instructions with many fields, such as the conditional compare of task D5\_02, bit-field moves and logical operations with an immediate. The third puts long runs of repeated bytes near the code to change, such as padding or tables of constants, which make the sequence to replace harder to copy exactly.

Two more kinds come from limits of the design. Without the padding after the function that Section~\ref{sec:threats} discusses, new code needs a code cave, a run of unused bytes elsewhere in the file, or a trampoline, a jump out to the new code and back, and the edit must also encode the offsets of the branches that reach it. Larger executables, the subject of Section~\ref{sec:scale}, and variable-length instructions, as on x86-64, are the other.

\subsection{Scale}
\label{sec:scale}

Larger files, the last kind of task above, raise a question of scale. With the executable given as bits, the prompt grows in proportion to the file, at 9 characters per byte. If the executable grows while the code to change stays the same size, that code sits among more bytes. In these executables the code section is 42 to 45 percent of the file, and with the bits only the failed tests of gpt-oss-120b and Qwen3-235B are mostly in the wrong place, as Section~\ref{sec:rq3} reports.

A large enough executable also exceeds the context of a model. A tool then has to select the part to give, for instance one function or one section located through the listing.

The curve of Supplementary Figure~S2, runs solved by the size of the edit, concerns the bytes that the reference solution changes. All executables in the study have about the same size. Whether the results carry over to larger executables is open; Section~\ref{sec:threats} states the scope of the study under external validity.

\subsection{Threats to validity}
\label{sec:threats}

\paragraph{Construct validity}

A run is solved when 5 to 8 hidden tests pass. An edit can pass them and still differ from the intended change elsewhere. Each task has tests of the old behaviour, and Figure~\ref{fig:routes} compares the bytes of each solved run with the compiler's. The attribution of a failed test to locating or writing uses a region, the function or data that the reference solution changes. A different region would move runs between wrong place and wrong change. It would not move the unsolved runs that fail before the tests.

The listing changes two things at once: the form in which the bytes are shown, and their decoding into instructions. The 6,000 runs cannot separate the two, so the decoding bottleneck as measured, the gain from the listing, includes the change of representation. A condition with a plain hexadecimal dump and no disassembly separates them. Section~\ref{sec:frontier} reports it for Claude Opus 5.5 at levels A and B, where six of the nine runs unsolved with the bits stay unsolved with hexadecimal. For the other five models the two effects are not separated.

In Supplementary Table~S1, the same prompt gives each model a different number of input tokens, because each model splits the bit sequence into tokens in its own way. This is part of the representation, and the study does not vary it.

In families V3 and D1, the change request quotes texts that the file stores as written, so the bytes to change can be found by searching for the text. In family V5, the quoted value is stored as a 4-byte integer, which must be encoded before it can be found, yet Supplementary Figure~S1 shows four models solving most of its runs with the bits only. In the instruction families, a quoted value is encoded inside an instruction.

\paragraph{Internal validity}
The runs use two providers, the default sampling parameters and one attempt per run. Claude Opus 5.5 was run through Anthropic on 22 November 2026, three days after the other models, and the control runs of Section~\ref{sec:frontier} on 24 November 2026; a higher effort level could change its few failures. The prompt, the requested output limit and the scoring are the same for every model and every condition. One attempt does not measure how much the answer to one prompt varies between calls; Section~\ref{sec:frontier} measures it for one model and one family. Each condition of a model has 100 runs on 100 different tasks, so the results are patterns across tasks.

Nemotron-3 Super is stopped at 32,768 output tokens, below the 60,000 of the other models. Figure~\ref{fig:budget} compares it with MiniMax-M3 at the same limit. The limit weighs most on insert logic: nearly all the insert logic runs that MiniMax-M3 solves use more than 32,768 output tokens, against few of its data only runs, and Nemotron-3 Super reaches its limit in almost every insert logic run, so its results on insert logic are bounded by it.

\paragraph{External validity}
The executables are for AArch64 only, are about 1.5 kilobytes and come from 20 families of generated programs, and six models from six developers were called through two providers. The results are measured on this architecture, this size and these models; Section~\ref{sec:scale} discusses larger executables. A single architecture keeps the encoding constant across the conditions, and a second one would double the runs.

The architecture also shapes the design. AArch64 instructions are all 4 bytes long, so replacing one never shifts the instructions after it; with variable-length instructions, as on x86-64, an edit in place has to pad or rearrange the code around it. Every reference solution overwrites bytes without changing the length of the file. In 34 of the 35 insert logic tasks it writes part of the new code into the padding that the alignment flag of Section~\ref{sec:build} leaves after the function. Without such padding, new code would need a code cave, a new section or a trampoline, and these are outside the design.

\paragraph{Conclusion validity}
The McNemar test treats the 500 pairs of a model as independent, although five pairs share each task. Section~\ref{sec:rq1} also compares, task by task, how often the listing raises and lowers the levels solved, and the tests of source information use the 100 tasks as blocks. In these data, small spreads between levels are not significant at 100 tasks per level. As a check, Supplementary Table~S2 fits a logistic regression per model that allows for the dependence between the runs of a task. It gives the same conclusion for the five models where it can be fitted: the listing raises the odds of solving for each of them, and source information matters only for DeepSeek-V4.1-Flash with the bits only. For Claude Opus 5.5 the regression cannot be fitted, because at some levels every run is solved; the McNemar and Cochran tests of Table~\ref{tab:tests} apply to it. Supplementary Tables~S5 and~S6 give the paired differences between levels with intervals and the comparisons of RQ2 with the family as the unit.

\section{Conclusion}
\label{sec:conclusion}

On these small AArch64 executables, a model can edit the bytes directly. For RQ1, the listing raises the runs solved for five of the six models, and one of them, DeepSeek-V4.1-Flash, solves nearly all its runs with it; the sixth, Claude Opus 5.5, solves nearly all its runs with the bits only. Source information changes the runs solved for these two models, in both cases with the bits only. For RQ2, the runs solved depend on the kind of change: with the bits only, five models solve data tasks far more often than instruction tasks, and with the listing, four models rarely solve tasks that insert new logic. For RQ3, most unsolved runs fail before the hidden tests. Among those that reach the tests, the two models with the most such runs fail mostly at locating with the bits only, and four models fail mostly at writing with the listing.

The results separate two bottlenecks. The first, the decoding bottleneck, is turning the bits into instructions: the listing reduces it for five models and nearly closes it for DeepSeek-V4.1-Flash. The second is producing a complete, applicable and correct edit once the instructions are given: with the listing, four models leave most runs unsolved. The failures of Claude Opus 5.5 sit in single bits of an encoding and point to the tasks that may stay hard, listed in Section~\ref{sec:hard}.

The results also suggest four elements of a method for direct editing, which Section~\ref{sec:method} states: the listing first, a byte-order check, edits measured against the compiler's bytes and an output limit set high enough. The elements come from observing every task of the study, so the next step is to test the method on tasks that the observation did not use, on larger executables and on the tasks that may stay hard.

\section*{CRediT authorship contribution statement}

\textbf{Camilo Chac\'on Sartori:} Conceptualization, Methodology, Software, Investigation, Formal analysis, Writing -- original draft, Writing -- review \& editing.

\section*{Declaration of competing interest}

The author declares no known competing financial interests or personal relationships that could have appeared to influence the work reported in this paper.

\section*{Funding}

This research did not receive any specific grant from funding agencies in the public, commercial, or not-for-profit sectors.

\section*{Data availability}

The tasks with their hidden tests, the executables and their disassembler listings, the 6,000 stored runs, the 250 control runs of Section~\ref{sec:frontier} and all scripts are deposited in a public archive, anonymised for review, at \url{https://doi.org/10.5281/zenodo.24449739}; the de-anonymised archive replaces it upon acceptance. The code is released under the MIT licence and the data under CC BY 4.0. Supplementary Figure~S3 shows the archive's structure, the map from each table and figure of the paper to the file and the script behind it, and the reproduction path: one command regenerates every result table and figure from the stored runs, offline and with no API key, byte-identical to the copies the archive ships. With Python 3 alone it regenerates all of them except Figure~\ref{fig:onebit} and Supplementary Tables~S2 to~S7; with its option \texttt{-{}-stats} it regenerates these too, which need four Python packages at versions the archive pins and a disassembler that reads AArch64.

\section*{Declaration of generative AI and AI-assisted technologies in the manuscript preparation process}

During the preparation of this work the author used Claude (Anthropic) to write the code that runs and analyses the experiments and draws the figures, and to polish the prose. After using this tool, the author reviewed and edited the content as needed and takes full responsibility for the content of the published article.

\bibliographystyle{elsarticle-num-names}
\bibliography{references}

\begin{thebibliography}{22}
\expandafter\ifx\csname natexlab\endcsname\relax\def\natexlab#1{#1}\fi
\providecommand{\url}[1]{\texttt{#1}}
\providecommand{\href}[2]{#2}
\providecommand{\path}[1]{#1}
\providecommand{\DOIprefix}{doi:}
\providecommand{\ArXivprefix}{arXiv:}
\providecommand{\URLprefix}{URL: }
\providecommand{\Pubmedprefix}{pmid:}
\providecommand{\doi}[1]{\href{http://dx.doi.org/#1}{\path{#1}}}
\providecommand{\Pubmed}[1]{\href{pmid:#1}{\path{#1}}}
\providecommand{\bibinfo}[2]{#2}
\ifx\xfnm\relax \def\xfnm[#1]{\unskip,\space#1}\fi
\bibitem[{Wenzl et~al.(2019)Wenzl, Merzdovnik, Ullrich, and
  Weippl}]{wenzl2019binaryrewriting}
\bibinfo{author}{M.~Wenzl}, \bibinfo{author}{G.~Merzdovnik},
  \bibinfo{author}{J.~Ullrich}, \bibinfo{author}{E.~Weippl},
\newblock \bibinfo{title}{{From Hack to Elaborate Technique---A Survey on
  Binary Rewriting}},
\newblock \bibinfo{journal}{ACM Computing Surveys} \bibinfo{volume}{52}
  (\bibinfo{year}{2019}) \bibinfo{pages}{1--37}.
  \DOIprefix\doi{10.1145/3316415}.
\bibitem[{Dai et~al.(2026)Dai, Priyadarshan, Imran, Wang, and
  Bianchi}]{dai2026scribe}
\bibinfo{author}{H.~Dai}, \bibinfo{author}{S.~Priyadarshan},
  \bibinfo{author}{A.~Imran}, \bibinfo{author}{R.~Wang},
  \bibinfo{author}{A.~Bianchi},
\newblock \bibinfo{title}{{Scribe: Practical Static Binary Patching via
  Binary-Aware Recompilation of Decompiled Code}},
\newblock in: \bibinfo{booktitle}{2026 IEEE 11th European Symposium on Security
  and Privacy (EuroS\&P)}, \bibinfo{publisher}{IEEE}, \bibinfo{year}{2026}, pp.
  \bibinfo{pages}{1218--1234}. \DOIprefix\doi{10.1109/eurosp68448.2026.00079}.
\bibitem[{Duck et~al.(2020)Duck, Gao, and Roychoudhury}]{duck2020e9patch}
\bibinfo{author}{G.~J. Duck}, \bibinfo{author}{X.~Gao},
  \bibinfo{author}{A.~Roychoudhury},
\newblock \bibinfo{title}{{Binary rewriting without control flow recovery}},
\newblock in: \bibinfo{booktitle}{Proceedings of the 41st ACM SIGPLAN
  Conference on Programming Language Design and Implementation},
  \bibinfo{publisher}{ACM}, \bibinfo{year}{2020}, pp.
  \bibinfo{pages}{151--163}. \DOIprefix\doi{10.1145/3385412.3385972}.
\bibitem[{Schulte et~al.(2022)Schulte, Brown, and Folts}]{schulte2022rewriters}
\bibinfo{author}{E.~Schulte}, \bibinfo{author}{M.~D. Brown},
  \bibinfo{author}{V.~Folts},
\newblock \bibinfo{title}{{A Broad Comparative Evaluation of x86-64 Binary
  Rewriters}},
\newblock in: \bibinfo{booktitle}{Proceedings of the 15th Workshop on Cyber
  Security Experimentation and Test}, \bibinfo{publisher}{ACM},
  \bibinfo{year}{2022}, pp. \bibinfo{pages}{129--144}.
  \DOIprefix\doi{10.1145/3546096.3546112}.
\bibitem[{Ma et~al.(2026)Ma, Du, Li, Meng, Fang, Ding, and
  Zheng}]{ma2026dlcodegen}
\bibinfo{author}{X.~Ma}, \bibinfo{author}{X.~Du}, \bibinfo{author}{C.~Li},
  \bibinfo{author}{J.~Meng}, \bibinfo{author}{X.~Fang},
  \bibinfo{author}{W.~Ding}, \bibinfo{author}{Z.~Zheng},
\newblock \bibinfo{title}{{How Do Large Language Models Perform in Deep
  Learning Code Generation? An Empirical Study}},
\newblock \bibinfo{journal}{ACM Transactions on Software Engineering and
  Methodology}  (\bibinfo{year}{2026}). \DOIprefix\doi{10.1145/3816024}.
\bibitem[{Vallecillos-Ruiz et~al.(2026)Vallecillos-Ruiz, Grishina, Hort, and
  Moonen}]{vallecillos2026roundtrip}
\bibinfo{author}{F.~Vallecillos-Ruiz}, \bibinfo{author}{A.~Grishina},
  \bibinfo{author}{M.~Hort}, \bibinfo{author}{L.~Moonen},
\newblock \bibinfo{title}{{Assessing the Latent Automated Program Repair
  Capabilities of Large Language Models Using Round-Trip Translation}},
\newblock \bibinfo{journal}{ACM Transactions on Software Engineering and
  Methodology} \bibinfo{volume}{35} (\bibinfo{year}{2026})
  \bibinfo{pages}{1--37}. \DOIprefix\doi{10.1145/3771922}.
\bibitem[{Abdollahi et~al.(2026)Abdollahi, Tasnia, Saha, Yang, Wang, and
  Hemmati}]{abdollahi2026reasoning}
\bibinfo{author}{M.~Abdollahi}, \bibinfo{author}{K.~R. Tasnia},
  \bibinfo{author}{S.~K. Saha}, \bibinfo{author}{J.~Yang},
  \bibinfo{author}{S.~Wang}, \bibinfo{author}{H.~Hemmati},
\newblock \bibinfo{title}{{Demystifying Errors in LLM Reasoning Traces: An
  Empirical Study of Code Execution Simulation}},
\newblock \bibinfo{journal}{ACM Transactions on Software Engineering and
  Methodology}  (\bibinfo{year}{2026}). \DOIprefix\doi{10.1145/3819239}.
\bibitem[{Shang et~al.(2025)Shang, Fu, Cheng, Chen, Li, Hu, Zhang, and
  Yu}]{shang2025binaryunderstanding}
\bibinfo{author}{X.~Shang}, \bibinfo{author}{Z.~Fu},
  \bibinfo{author}{S.~Cheng}, \bibinfo{author}{G.~Chen},
  \bibinfo{author}{G.~Li}, \bibinfo{author}{L.~Hu}, \bibinfo{author}{W.~Zhang},
  \bibinfo{author}{N.~Yu},
\newblock \bibinfo{title}{{An empirical study on the effectiveness of large
  language models for binary code understanding}},
\newblock \bibinfo{journal}{Empirical Software Engineering}
  \bibinfo{volume}{31} (\bibinfo{year}{2025}).
  \DOIprefix\doi{10.1007/s10664-025-10748-5}.
\bibitem[{Li et~al.(2026)Li, Li, Gao, Gao, Li, Xiao, Li, and
  Luo}]{li2026binaryspd}
\bibinfo{author}{Q.~Li}, \bibinfo{author}{B.~Li}, \bibinfo{author}{C.~Gao},
  \bibinfo{author}{S.~Gao}, \bibinfo{author}{Z.~Li}, \bibinfo{author}{Y.~Xiao},
  \bibinfo{author}{C.~Li}, \bibinfo{author}{B.~Luo},
\newblock \bibinfo{title}{{Empirical Study of Code Large Language Models for
  Binary Security Patch Detection}},
\newblock \bibinfo{journal}{ACM Transactions on Software Engineering and
  Methodology}  (\bibinfo{year}{2026}). \DOIprefix\doi{10.1145/3822506}.
\bibitem[{Tan et~al.(2024)Tan, Luo, Li, and Zhang}]{tan2024llm4decompile}
\bibinfo{author}{H.~Tan}, \bibinfo{author}{Q.~Luo}, \bibinfo{author}{J.~Li},
  \bibinfo{author}{Y.~Zhang},
\newblock \bibinfo{title}{{LLM4Decompile: Decompiling Binary Code with Large
  Language Models}},
\newblock in: \bibinfo{booktitle}{Proceedings of the 2024 Conference on
  Empirical Methods in Natural Language Processing},
  \bibinfo{publisher}{Association for Computational Linguistics},
  \bibinfo{year}{2024}, pp. \bibinfo{pages}{3473--3487}.
  \DOIprefix\doi{10.18653/v1/2024.emnlp-main.203}.
\bibitem[{Cui et~al.(2026)Cui, Gao, Qin, Wang, Wang, Luo, Zhou, and
  Zhang}]{cui2026pcodetrans}
\bibinfo{author}{Y.~Cui}, \bibinfo{author}{Z.~Gao}, \bibinfo{author}{S.~Qin},
  \bibinfo{author}{Y.~Wang}, \bibinfo{author}{J.~Wang},
  \bibinfo{author}{X.~Luo}, \bibinfo{author}{P.~Zhou},
  \bibinfo{author}{C.~Zhang},
\newblock \bibinfo{title}{{PCodeTrans: Translate Decompiled Pseudocode to
  Trace-Level Equivalent Source Code}},
\newblock in: \bibinfo{booktitle}{Proceedings of the 41st IEEE/ACM
  International Conference on Automated Software Engineering},
  \bibinfo{publisher}{ACM}, \bibinfo{year}{2026}, pp.
  \bibinfo{pages}{1715--1727}. \DOIprefix\doi{10.1145/3832783.3837432}.
\bibitem[{Cummins et~al.(2025)Cummins, Seeker, Grubisic, Roziere, Gehring,
  Synnaeve, and Leather}]{cummins2025llmcompiler}
\bibinfo{author}{C.~Cummins}, \bibinfo{author}{V.~Seeker},
  \bibinfo{author}{D.~Grubisic}, \bibinfo{author}{B.~Roziere},
  \bibinfo{author}{J.~Gehring}, \bibinfo{author}{G.~Synnaeve},
  \bibinfo{author}{H.~Leather},
\newblock \bibinfo{title}{{LLM Compiler: Foundation Language Models for
  Compiler Optimization}},
\newblock in: \bibinfo{booktitle}{Proceedings of the 34th ACM SIGPLAN
  International Conference on Compiler Construction}, \bibinfo{publisher}{ACM},
  \bibinfo{year}{2025}, pp. \bibinfo{pages}{141--153}.
  \DOIprefix\doi{10.1145/3708493.3712691}.
\bibitem[{Shang et~al.(2025)Shang, Chen, Cheng, Wu, Hu, Li, Zhang, and
  Yu}]{shang2025binmetric}
\bibinfo{author}{X.~Shang}, \bibinfo{author}{G.~Chen},
  \bibinfo{author}{S.~Cheng}, \bibinfo{author}{B.~Wu}, \bibinfo{author}{L.~Hu},
  \bibinfo{author}{G.~Li}, \bibinfo{author}{W.~Zhang}, \bibinfo{author}{N.~Yu},
\newblock \bibinfo{title}{{BinMetric: A Comprehensive Binary Code Analysis
  Benchmark for Large Language Models}},
\newblock in: \bibinfo{booktitle}{Proceedings of the Thirty-Fourth
  International Joint Conference on Artificial Intelligence},
  \bibinfo{publisher}{International Joint Conferences on Artificial
  Intelligence Organization}, \bibinfo{year}{2025}, pp.
  \bibinfo{pages}{7715--7723}. \DOIprefix\doi{10.24963/ijcai.2025/858}.
\bibitem[{Xie et~al.(2024)Xie, Zhang, Jiang, Xu, Tan, and Zhang}]{xie2024resym}
\bibinfo{author}{D.~Xie}, \bibinfo{author}{Z.~Zhang},
  \bibinfo{author}{N.~Jiang}, \bibinfo{author}{X.~Xu},
  \bibinfo{author}{L.~Tan}, \bibinfo{author}{X.~Zhang},
\newblock \bibinfo{title}{{ReSym: Harnessing LLMs to Recover Variable and Data
  Structure Symbols from Stripped Binaries}},
\newblock in: \bibinfo{booktitle}{Proceedings of the 2024 on ACM SIGSAC
  Conference on Computer and Communications Security},
  \bibinfo{publisher}{ACM}, \bibinfo{year}{2024}, pp.
  \bibinfo{pages}{4554--4568}. \DOIprefix\doi{10.1145/3658644.3670340}.
\bibitem[{Jiang et~al.(2025)Jiang, Wang, Liu, Xu, Tan, Zhang, and
  Babkin}]{jiang2025nova}
\bibinfo{author}{N.~Jiang}, \bibinfo{author}{C.~Wang},
  \bibinfo{author}{K.~Liu}, \bibinfo{author}{X.~Xu}, \bibinfo{author}{L.~Tan},
  \bibinfo{author}{X.~Zhang}, \bibinfo{author}{P.~Babkin},
\newblock \bibinfo{title}{{Nova: Generative Language Models for Assembly Code
  with Hierarchical Attention and Contrastive Learning}},
\newblock in: \bibinfo{booktitle}{International Conference on Learning
  Representations (ICLR 2025)}, \bibinfo{year}{2025}. \URLprefix
  \url{https://openreview.net/forum?id=4ytRL3HJrq}.
\bibitem[{Wei et~al.(2026)Wei, Suresh, Tan, Xu, Singh, Wang, and
  Aiken}]{wei2026supercoder}
\bibinfo{author}{A.~Wei}, \bibinfo{author}{T.~Suresh},
  \bibinfo{author}{H.~Tan}, \bibinfo{author}{Y.~Xu},
  \bibinfo{author}{G.~Singh}, \bibinfo{author}{K.~Wang},
  \bibinfo{author}{A.~Aiken},
\newblock \bibinfo{title}{{SuperCoder: Assembly Program Superoptimization with
  Large Language Models}},
\newblock in: \bibinfo{booktitle}{Third Conference on Language Modeling},
  \bibinfo{year}{2026}. \URLprefix
  \url{https://openreview.net/forum?id=VsaeinNuM7}.
\bibitem[{Wu et~al.(2024)Wu, Tan, Wang, Wang, Li, and Sun}]{wu2024bgpt}
\bibinfo{author}{S.~Wu}, \bibinfo{author}{X.~Tan}, \bibinfo{author}{Z.~Wang},
  \bibinfo{author}{R.~Wang}, \bibinfo{author}{X.~Li}, \bibinfo{author}{M.~Sun},
  \bibinfo{title}{{Beyond Language Models: Byte Models are Digital World
  Simulators}}, \bibinfo{howpublished}{arXiv preprint arXiv:2402.19155},
  \bibinfo{year}{2024}. \DOIprefix\doi{10.48550/arXiv.2402.19155}.
\bibitem[{St{\"o}rtz et~al.(2026)St{\"o}rtz, Stan, Dinu, Servia-Rodr{\'i}guez,
  Gaman, Miron, and Raff}]{stortz2026lbm}
\bibinfo{author}{F.~St{\"o}rtz}, \bibinfo{author}{C.-A. Stan},
  \bibinfo{author}{A.~Dinu}, \bibinfo{author}{S.~Servia-Rodr{\'i}guez},
  \bibinfo{author}{M.~Gaman}, \bibinfo{author}{C.~Miron},
  \bibinfo{author}{E.~Raff}, \bibinfo{title}{{Large Byte Model: Teaching
  Language Models About Compiled Code}}, \bibinfo{howpublished}{arXiv preprint
  arXiv:2606.02834}, \bibinfo{year}{2026}.
  \DOIprefix\doi{10.48550/arXiv.2606.02834}.
\bibitem[{McNemar(1947)}]{mcnemar1947}
\bibinfo{author}{Q.~McNemar},
\newblock \bibinfo{title}{{Note on the Sampling Error of the Difference Between
  Correlated Proportions or Percentages}},
\newblock \bibinfo{journal}{Psychometrika} \bibinfo{volume}{12}
  (\bibinfo{year}{1947}) \bibinfo{pages}{153--157}.
  \DOIprefix\doi{10.1007/BF02295996}.
\bibitem[{Cochran(1950)}]{cochran1950}
\bibinfo{author}{W.~G. Cochran},
\newblock \bibinfo{title}{{The Comparison of Percentages in Matched Samples}},
\newblock \bibinfo{journal}{Biometrika} \bibinfo{volume}{37}
  (\bibinfo{year}{1950}) \bibinfo{pages}{256--266}.
  \DOIprefix\doi{10.1093/biomet/37.3-4.256}.
\bibitem[{Holm(1979)}]{holm1979}
\bibinfo{author}{S.~Holm},
\newblock \bibinfo{title}{{A Simple Sequentially Rejective Multiple Test
  Procedure}},
\newblock \bibinfo{journal}{Scandinavian Journal of Statistics}
  \bibinfo{volume}{6} (\bibinfo{year}{1979}) \bibinfo{pages}{65--70}.
\bibitem[{Li and Paxson(2017)}]{li2017securitypatches}
\bibinfo{author}{F.~Li}, \bibinfo{author}{V.~Paxson},
\newblock \bibinfo{title}{A large-scale empirical study of security patches},
\newblock in: \bibinfo{booktitle}{Proceedings of the 2017 ACM SIGSAC Conference
  on Computer and Communications Security}, \bibinfo{publisher}{ACM},
  \bibinfo{year}{2017}, pp. \bibinfo{pages}{2201--2215}.
  \DOIprefix\doi{10.1145/3133956.3134072}.

\end{thebibliography}

\end{document}